\documentclass[%
 reprint,
 amsmath,amssymb,
 aps,
pre,
floatfix,
]{revtex4-1}

\usepackage{graphicx}
\usepackage{dcolumn}
\usepackage{bm}
\usepackage{epstopdf}
\usepackage{siunitx}
\usepackage{gensymb}

\begin{document}


\title{Dynamics of Spiral Spin Waves in Magnetic Nano-patches: Influence of Thickness and Shape}

\author{D. Osuna Ruiz *\textsuperscript{1}} 
\email{do278@exeter.ac.uk}
\author{E. Burgos Parra\textsuperscript{1}}
\author{N. Bukin\textsuperscript{1}}
\author{M. Heath\textsuperscript{1}}
\author{A. Lara\textsuperscript{2}}
\author{F. G. Aliev\textsuperscript{2}}
\author{A. P. Hibbins\textsuperscript{1}}
\author{F. Y. Ogrin\textsuperscript{1}}%
 
\affiliation{%
 \textsuperscript{1}Department of Physics and Astronomy, University of Exeter, Exeter EX4 4QL, United Kingdom.
 }%
\affiliation{%
\textsuperscript{2}Departamento de Fisica de la Materia Condensada, IFIMAC and INC, Universidad Autonoma de Madrid, Cantoblanco 28049, Madrid, Spain
}%

\date{\today}

\begin{abstract}
We explore the dynamics of spiral spin waves in Permalloy nano-elements with variable aspect ratio of geometric dimensions, and their potential use as improved spin wave emitters with no or little biasing field required. Numerical results show that above a certain thickness, propagating spiral waves can be obtained in circular and square shaped elements in a flux closure state. VNA-FMR experiments on 20 nm (thin) and 80 nm (thick) samples confirm two type of spectra corresponding to different dispersions for thinner and thicker elements.  We show that, for the thicker films, the vortex core region acts as a source of large amplitude spiral spin waves, which dominate over other modes. In case of the thinner elements, these modes are critically damped. For different shapes of the patch, we show that a rich collection of confined propagating modes can also be excited, modifying the final wave front and enriching the potential of the nano-dot as a spin wave emitter.  We give an explanation for the intense spiral modes from the perspective of a balance of dipolar and exchange energies in the sample. 

\end{abstract}

\pacs{Valid PACS appear here}
\maketitle


\section{\label{sec:level1}introduction}

Spin waves in confined structures have been a subject of interest with the view of their potential applications in the field of computing and data storage \cite{0022-3727-43-26-264001,Karenowska2016,PhysRevApplied.4.047001,0953-8984-26-12-123202}. Due to a competition between dipolar and exchange energies in such structures, a range of relaxed states can be achieved, dependent on the shape and the dimensions of the structures. Most of these magnetization states will have a preferred in-plane magnetization with inhomogeneities arising at the boundaries or high symmetry points, leading to magnetization singularities such as magnetic domain walls and vortex configurations. It has been reported that, due to the confinement or the natural magnetic state of the sample, inhomogeneities of the internal magnetic field can be sources of spin waves due to a graded index in the magnonic landscape \cite{PhysRevB.96.064415, article}. A number of studies have focused on characterizing and explaining the source of spin waves in confined micron and sub-micron structures for different magnetic domain configurations and saturated states \cite{PhysRevB.76.224401, PhysRevB.73.104424,articlesquareseigenmodes}. Due to the finite size of the structures, dispersion relations of spin waves are expected to be discrete in contrast to a continuum spectra for spin waves in infinite films \cite{doi:10.1063/1.1713058}. Spin wave propagation along the edges and emission from edges and local regions in the magnet have also been proved through micromagnetic simulations and direct imaging \cite{PhysRevB.96.094430, doi:10.1063/1.4995991,propedges}. Moreover, in circular disks, depending on the orientation of the oscillating field, i.e. transverse or normal to the plane, a spiral or circular standing pattern can be formed, respectively.

In some applications based on confined magnetic structures like nano-dots, the existence of spin waves is usually a non-desirable phenomena, as their frequencies lay in the GHz range and potentially interfere with writing or reading operations in magnetic memories. On the other hand, spin waves may play a central role in future communication technologies. For example, previous works have reported conceptual ideas and realizations of several magnonic devices, logic gates \cite{doi:10.1063/1.2089147,doi:10.1063/1.2975235} and information processing components such as transistors or diodes \cite{Chumak2014MagnonTF,PhysRevX.5.041049} which, in addition to an effective control on redirection along magnetic tracks \cite{doi:10.1063/1.4738887,articlebending}, can lead to more complex and practical magnonic logic circuitry. Hence, discovering new ways to effectively source and control spin wave propagation can make a significant impact on realization of such applications.

In this paper we focus on the problem of spin-wave emission of thick magnetic elements with vortex singularities. Wintz et al. \cite{articlewintz} demonstrated recently that a vortex core can serve as a source of spin wave emission in ferromagnetic bilayer elements, with the ability of tuning the spin wave frequency and propagation direction, inwards towards, or outwards from the core. They reported large amplitude spin wave modes with spiral wavefront. A further work by the same team has also reported spiral spin waves in monolayered nano-disks, which were explained as a result of  hybridization with first order standing spin waves across the thickness of the disk \cite{2017arXiv171200681D}. 
The understanding of the origin of these modes is, by itself, an interesting topic and consequently led to a number of subsequent studies, for example by Kammerer et al. \cite{Kammerer2011MagneticVC} and Stoll et al. \cite{10.3389/fphy.2015.00026}. However these studies mainly focused on stationary modes and core reversal, with the formation of a dynamic double core or \lq dip'. A similar explanation was offered by Verba et al. \cite{PhysRevB.93.214437}, but through a different approach: varying thickness to compare different hybridizations in thin and thick samples. 

Our present study explores the formation of spiral-spin waves in thicker samples using both numerical simulations and experimental results. Starting from the principle formalism of dynamics described by the Landau-Lifshitz equation of motion we consider the different energy contributions and their influence in thin and thick elements. We consider the dispersion of the supported spin-wave, and study the influence of thickness on the intensity enhancement of these spiral modes. We also provide insight into the way that the element shape affects the magnetization distribution by making comparisons between a square sample and the original circular one. We study the arising of confined modes in different types of patches, which can lead to further enhancement of the spiral wave and turning the magnetic patch into a more reliable and complex spin wave emitter. We experimentally characterize these spin wave modes using a standard VNA-FMR (\lq Vector Network Analyzer - Ferromagnetic Resonance') setup, in contrast to the more complex imaging techniques such as space resolved Kerr microscopy, used by Neudecker et al. \cite{PhysRevB.73.134426}. Our findings may lead to the development of new types of efficient spin wave emitters for information technologies, potentially useful as simple, tunable and high-intensity source units in larger magnonic circuitry.

In section II, we describe our numerical model, and in section III, we show our experimental and numerical results, including evidence illustrating the spiral character of the mode propagation and its dispersion relations (for both circular and square geometries). 

\section{\label{sec:level2}Materials and methods}

In order to understand the dynamics of spiral spin waves, a set of micromagnetic simulations using Mumax3 software \cite{doi:10.1063/1.4899186} have been undertaken. The typical material parameters of Permalloy elements at room temperature were \cite{doi:10.1063/1.4899186} : saturation magnetization $M_s$ $= 8$ $\times$ $10^5$ Am$^{-1}$, exchange constant $A_{ex} = $ $1.3$ $\times$  $10^{-11}$ J/m and Gilbert damping constant $\alpha$ $=$ $0.008$. All simulations, as well as the experimental conditions, assume room temperature parameters. Simulations were performed using a hexaedral grid for the samples with the following geometric parameters of the elements and the corresponding cell sizes: Circular (square) samples have a diameter ${d}$ (side length ${L}$) of 900 nm and different thicknesses ${t}$ of 20 and 80 nm. With a fixed size cell along {z} ({$N_{Cz}$}) of 5 nm for the thick samples and 2 nm for the thin, the grid is discretized in the ${x, y, z}$ space into 256 $\times$ 256 $\times$ $ {t} /{N_{Cz}}$ cells. Therefore, the number of cells in ${z}$ is 10 (${t} =$ 20 nm) and 16 (${t =}$ 80 nm). Cell size along {x} and {y} is  3.5 nm. Size cell along three dimensions is always kept smaller than the exchange length of Permalloy (5.7 nm). To verify that the cell size is not effecting the results, several simulations were performed with a reduced dimensions keeping the size of the elements fixed. Following the common practice \cite{doi:10.1063/1.4899186},  the number of cells were chosen to be powers of 2 to maintain the computational efficiency. In circular disks, we also set a \lq smooth edges' condition with value 8.
A key point for most micromagnetic simulations is to achieve a stable magnetization ground state. This was achieved by firstly setting an artificial vortex state with polarity and chirality numbers of  (1, $-$1) and then running the simulation with high damping term ($\alpha$ $=$ 1) to relax the magnetization until the maximum torque (\lq maxTorque' parameter in Mumax3) is at the level of {$10^{-7}$} T. Numerical value oscillates around a specific value, which indicates convergence and the effective achievement of a magnetization ground state. A typical time to achieve this is of the order of 100 ns. Once the ground state is obtained the whole magnetization configuration is stored and then used for the simulations with the dynamic activation.
To achieve an equal excitation across a desired frequency range in the dispersion diagram, a sinc-shaped magnetic pulse has been used,  ($B_{1}$(t)):

\begin{gather}
B_{1}(t) = A_{1}\text{sinc}(2\pi f_{c}( t-t_{d} )), 
\end{gather}

where {$f_{c}$} is the cut-off frequency, set to 30 GHz, and {$A_{1}$ = 10 mT} is the pulse amplitude. Using this activation, each mode is equivalently fed with an AC magnetic field of 0.3 mT in amplitude. This field is small enough to remain in the linear regime of activation and to avoid any static changes in the magnetic domain structure of the samples. The delay time {$t_{d}$ = 5 ns} provides a reasonable offset to the peak of the pulse, allowing a gradual increase of the amplitude from the beginning of the simulation. In another scenario, for analyzing time evolution of the magnetic signal, we apply a continuous wave (CW) excitation with a magnetic field $B_2$ at a specific frequency {$f_{0}$}:

\begin{gather}
B_{2}(t) = A_{2}\text{sin}(2\pi f_{0} t)), 
\end{gather}

Similar to the sinc pulse, each mode is excited with a relatively small oscillating field. Equivalently to the previous case, {$A_{2}$} is chosen to be 0.3 mT to obtain a good magnetic contrast in the time domain. 
\par We use a sampling period of {$T_{s} = $25 ps}  and record up to 1024 frames in space and time. With these parameters, simulation time is long enough (up to 50 ns) to get a reasonably good Fourier resolution of 78 MHz per frequency bin. For computational efficiency, we use a power of 2 for the number of time samples for faster Fourier transformation in analysis. The simulated spin wave spectra is calculated from the absolute amplitude of all modes averaged for every cell in the model. 
\par In the experiment, these spectra are compared to the absorption intensity, visualized through the real part of magnetic susceptibility ($\chi'$). Experimental results were obtained by measuring a transmission spectra of a sample in a coplanar waveguide geometry by means of a Vector Network Analyzer ferromagnetic resonance (VNA-FMR) setup. We use wide frequency band VNA-FMR technique in a lower range of frequencies, where many waveguides operate with higher reliability. In our measurements we used a set of samples fabricated using standard EBL techniques, with different parameters of shapes and dimensions. To enhance the RF response the elements were grouped in square lattices with the separation equal to the size of the element, assuming no or very limited interaction between the elements. We tested arrays of  thick ({t} = 80 nm) disks and squares with diameter and side lengths ranging from 500 nm to 900 nm. Since the arrays of samples are placed on top of a Coplanar Waveguide (CPW), the pumping magnetic field provided by the CPW is applied parallel to the sample plane and, in the case of square elements, parallel to one side of the square lattice.  The sample is placed on the short path of a U-shaped CPW, and so, a perpendicular pumping with respect to the biasing field is applied ($H_{bias}{\bot }  h_{rf}$), \cite{PhysRevB.79.174433}. The experiments were carried out in the range of -70 mT to 70 mT, which was sufficient to cover the vortex structure regime in the elements as well as saturation, in steps of 2 mT. Each spectrum was reduced by subtracting a signal produced at a highest field, and then averaged over 10 measurements. This helped to remove the larger background signal related to the transmission line noise, and allowed us to amplify the magnetic response which is normally much smaller than other electromagnetic components.

\section{Numerical results and discussion}

In this section we explore spiral spin waves propagating outwards from the vortex core region. We analyze their dispersion characteristics for circular and square disks and provide an explanation from the point of view of dipolar and exchange energies and their balance.

\subsection{Dispersion Characterization} 
We start with the analysis of circular disks. Spin wave spectra of circular nanodots has been studied intensively in the past \cite{PhysRevB.93.184427, doi:10.1063/1.1772868}. When applying an in-plane excitation, the lowest state mode that can be observed is related to a gyrotropic motion of the vortex core, with the frequency of gyration highly dependent on the aspect ratio of the dot \cite{YuCore}. With increase of frequency, higher order gyrotropic modes \cite{higher, Guslienko2015GiantMV} and a complete set of modes related to azimuthal and radial spin waves can be obtained \cite{PhysRevB.93.184427}. The last type of spin waves are intuitively related to the well-known Damon-Eshbach modes in magnetized thin films, in the sense that radial $\mathbf{k}$ vectors are always perpendicular to $\mathbf{M}$ in a vortex core configuration \cite{PhysRev.118.1208}. Depending on the thickness and more generally, on physical dimensions, the spectra of these waves can drastically change. 

\par In our study, we firstly proceed to obtain dispersion relations for Permalloy circular disks with a relatively high aspect ratio (see Materials and Methods). To understand their dynamic properties here we explore their dispersion relations. Because of the circular geometry, which does not allow to take advantage of periodic boundary conditions (PBC), we have also built rectangular \lq semi-infinite' stripes, setting an equivalence between the radius of the disk and the lateral finite dimension (x axis) of the rectangular stripe. PBC along y axis of the stripe structure are set, this provides a similar dependence but in a much clear representation. Their magnetization distribution along x-direction is the same as in the circular disk whereas an infinite Bloch wall, similar to the one studied in \cite{PhysRevB.96.064415}, plays an equivalent role to that of the vortex core in disks. 
\par Figure 1 (a) shows a comparison of the simulated spin wave spectra for  the above mentioned structures: a 20 nm and 80 nm thick disks as well as the semi-infinite rectangles of the same thickness. Fig. 1(b) and (c) show their dispersion relations (for thick and thin structures, respectively). Although both 80 nm thick systems show more intense modes above 6 GHz than the thin elements, there are some small discrepancies between the resonance spectrum of the circular disk and the stripe, which could be due to the differences in shape anisotropy and the implications on the internal effective field and the local FMR \cite{doi:10.1063/1.4995991}. However, the difference in frequency spacing between relative larger peaks when the size is reduced (from d = 900 nm to 600 nm) remains the same ({$\Delta{f}$} = 0.77 GHz) for the disk and our stripe model, which suggests that this mode is of the same nature in both shapes and equally influenced by the cavity length of the shape.

\begin{figure}[ht]
\centering 
\includegraphics[trim=0cm 0cm 0cm 0cm, clip=true, width=8cm]{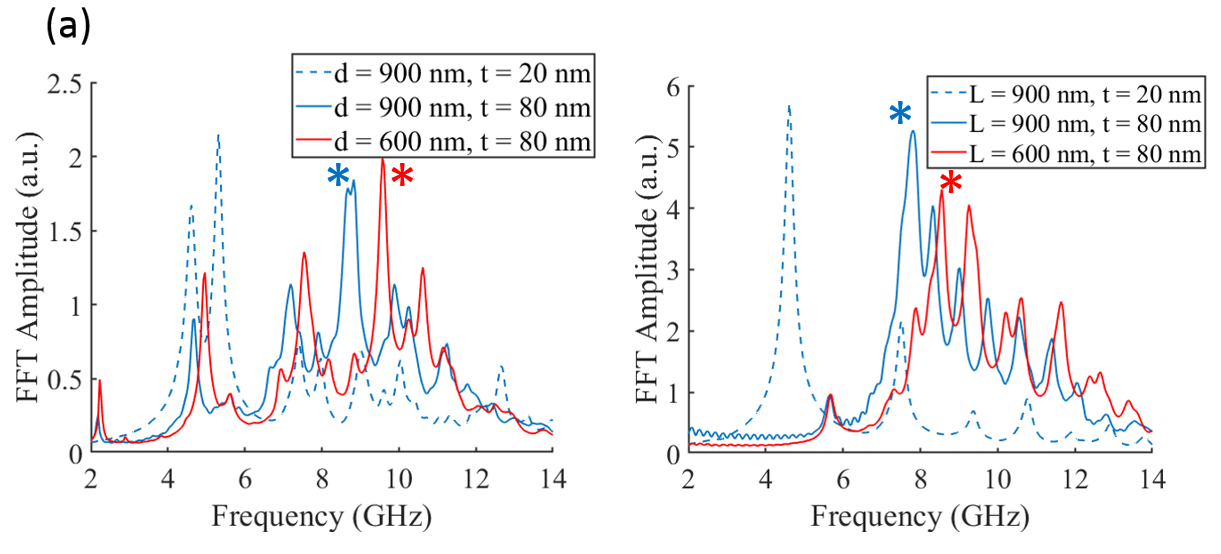}
\includegraphics[trim=0cm 0cm 0cm 0cm, clip=true, width=8cm]{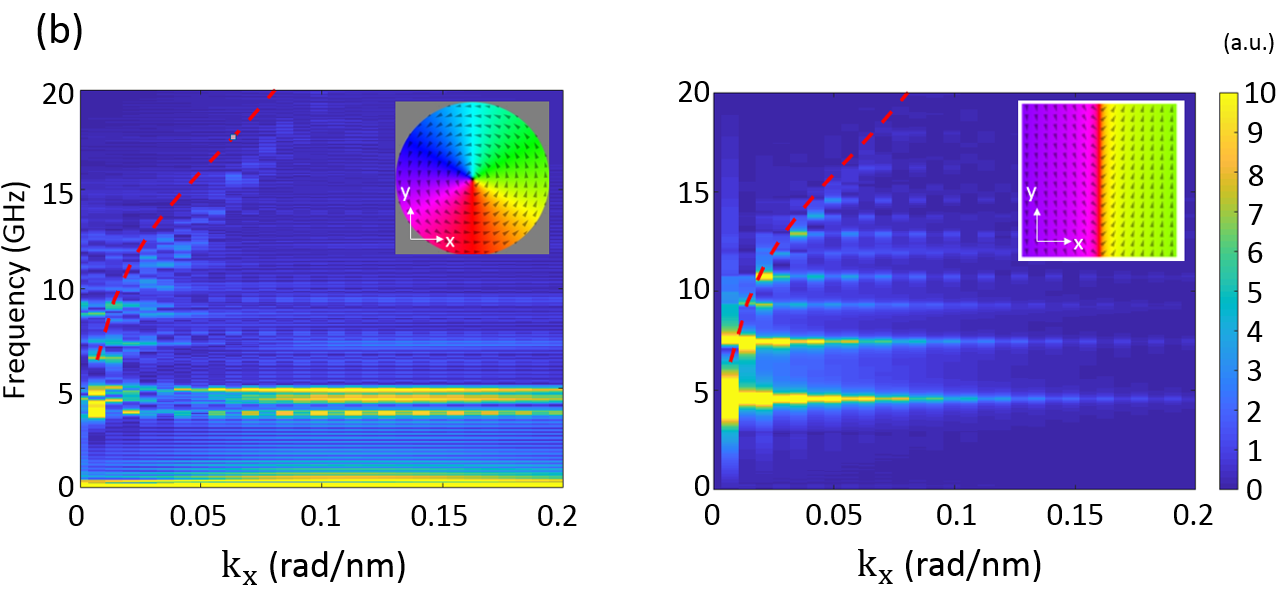}
\includegraphics[trim=0cm 0cm 0cm 0cm, clip=true, width=8cm]{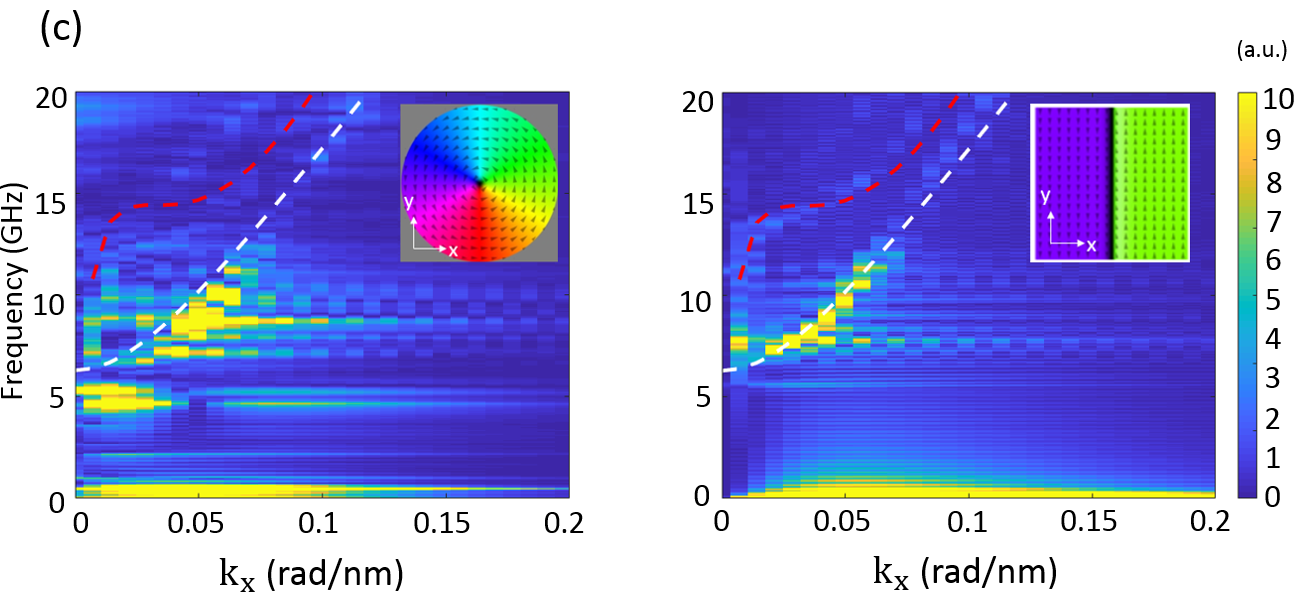}
\caption{(a) Spin wave spectra for thick and thin circular disks (left) and semi-infinite stripes (right). The main peak has been marked with a star. Dispersion diagrams in {x} direction, i.e., across the center for thin (b) and thick (c) disks and stripes of equal length along {x} (L = 900 nm) are shown. Color scale shows the module of each mode. The change in width of the vortex core can be easily spotted in the insets. Red dash line shows the analytical equation for a Damon-Eshbach mode from \cite{BRACHER20171} and white dashed line the analytical model from \cite{Kalinikos_1986} assuming unpinned moments.}  \label{Fields}
\end{figure}

\par It is also worth to note that the increase of thickness affects the magnetic inhomogeneity at the centre of the disk or the stripe, widening the vortex core and turning the Neel wall into a Bloch wall, respectively (see insets in Fig. 1(b) and (c)), being this the most prominent change in the magnetic configuration. All this suggests a connection with the enhanced mode above 6 GHz, which is not excited from the edges but rather, from the central region of the shape, i.e., from the vicinity of the vortex core (see Fig. 2) and, equivalently, from the vicinity of the Bloch wall in the semi-infinite stripe. The latter is supported by results from \cite{PhysRevB.96.064415}, suggesting that the inhomogeneity at the core region can be treated as an equivalent source of spin wave. For thicker disks, this implies that there is a multi-mode regime above a certain frequency, where the spin waves are being excited from both the edge and from the core, like it was found in \cite{doi:10.1063/1.4995991}. However, in our case, the waves originating from the core significantly prevail in amplitude. The obtained dispersion relations,  help to understand the spectra in more detail, see Fig. 1. For thin (20 nm) disks (Fig. 1(b)), we can clearly see that the only propagating mode excited at frequencies above 5 GHz is originating from the edges (see Fig. 2(a)), in accordance with the analytical model for spin waves in a Damon-Eshbach configuration from \cite{BRACHER20171}, which takes into account the finiteness of the element.  This is also consistent with previous results where the edges act as spin wave emitters \cite{doi:10.1063/1.4995991}.  In thicker disks (Fig. 1(c)), the lower branch mode can be analytically described assuming unpinned magnetic moments conditions, i.e., free surface boundary conditions for first higher order modes \cite{Kalinikos_1986,2017arXiv171200681D} (see Fig. 1).

\begin{figure}[ht]
\centering 
\includegraphics[trim=0cm 0cm 0cm 0cm, clip=true, width=8cm]{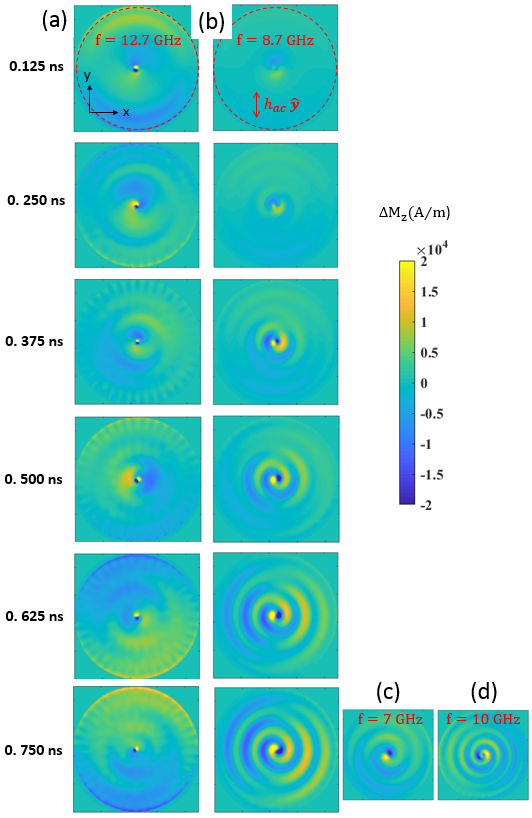}
\caption{Snapshots of the dynamic out-of-plane component of magnetization for thin circular disks, d = 900 nm, applying a CW in-plane excitation of 12.7 GHz (a) and for thick circular disks applying 8.7 GHz (b), 7 GHz (c) and 10 GHz (d) to the whole shape. While in thin disks the excited wave propagates inwards (a), on the thick disk the spin wave propagates outwards with more intensity (b).}  \label{Fields}
\textbf{}\end{figure}

\begin{figure}[ht]
\centering 
\includegraphics[trim=0cm 0cm 0cm 0cm, clip=true, width=8cm]{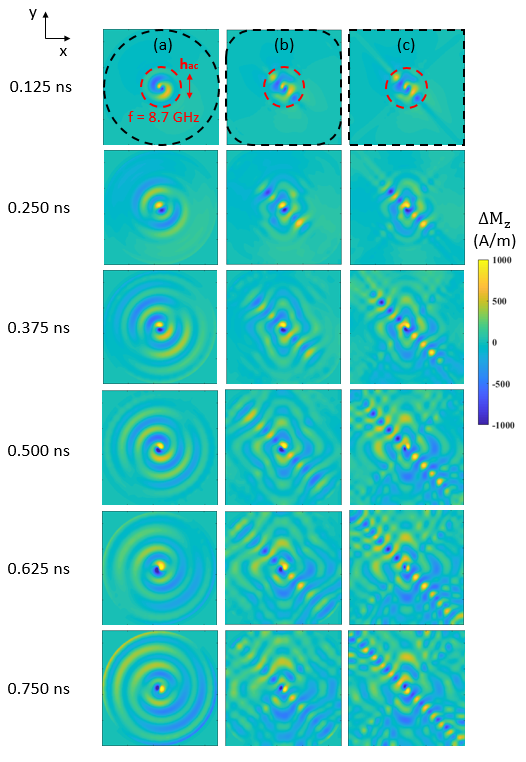}
\caption{Snapshots of the dynamic out-of-plane component of magnetization for thick circular disks ({d} = 900 nm) (a), rounded squared disks (b) and squared disks (c) applying a 8.7 GHz CW excitation at the core region ({d} = 100 nm). Black dashed contours for each shape are plotted and curvature of the chamfered corner is equal to one quarter of the side length.}  \label{Fields}
\end{figure}

\begin{figure}[ht]
\centering 
\includegraphics[trim=0cm 0cm 0cm 0cm, clip=true, width=7cm]
{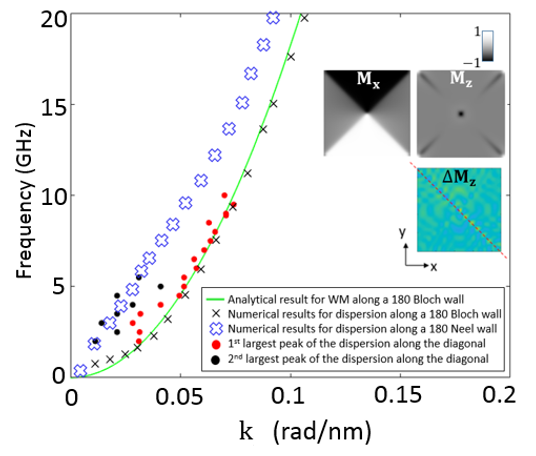}
\caption{Dispersion relation from of the mode traveling along the domain wall in the square (red dots). Numerical results from the dispersion along the diagonal of the square (red dots) agree nicely with the theoretical equation of a Winter's magnon along a 180 degrees Bloch wall (green solid line) in the exchange regime \cite{PhysRevLett.114.247206} at large k and show a mixing of two behaviors with the dispersion along a Neel wall at lower k and bloch at larger k. Insets show components of static magnetization demonstrating the
”mixed” Bloch and Neel wall behavior along the diagonal of the square.}  \label{Fields}
\end{figure}

\par Here we reproduce some results for thin circular disks for the comparison with our thicker samples. As previously reported for this case, an in-plane CW excitation gives rise to an inwards propagating wave which generates a spiraling wavefront. It is worth to note that this case is realized for the excitation in-plane, in comparison to radial waves which can be also produced with out-of-plane pumping \cite{doi:10.1063/1.4995991}. 
\par Looking at the thicker 80 nm disks we can clearly see the following factors in Fig. 2(b). If we excite a wave with an in-plane CW magnetic signal at the resonance condition of 8.7 GHz, we can observe the formation of a spiral pattern which is clearly emanating from the vortex core region of the disk. The spin wave's wavefront is attenuated along the direction of excitation $\mathbf{y}$,  that we believe is due to the destructive interference with the weaker oscillations coming from the edges, which are also excited along that direction. As expected from the dispersion relations, we get similar spiral patterns at different CW frequencies that fall into the excitation range of the lower branch (see Fig.2(c) and (d)). Also, the wavelength of these waves can be easily tuned changing the frequency of the driving field, similarly as it was reported in \cite{articlewintz}. Further simulations (not shown here) show that if the pumping field is applied out of plane, the outward propagating spiral wave turns into a circular wave. At higher frequencies, waves from the core become weaker and spin waves from the edges are now more efficiently excited. Another important feature of the spiral wave is that it is still excited at the resonance frequency even if the vortex core is shifted by applying a biasing external field, where the limit is saturation field.

\par Similar investigation was carried out in 80 nm thick Permalloy square elements. This shape shows a richer collection of modes due to the formation of domain walls, which provide additional topological confinement. In particular, they can serve as the channels for spin wave propagation, when the sample's magnetization configuration is in its ground state of closed leaf, Landau pattern \cite{articlesquareseigenmodes, Wagner2016MagneticDW}. Due to the Landau pattern, separation between the triangular domains in the square dot will give rise to 90 degrees domain walls. From this point, every mention in the article to \lq domain walls' in the studied samples or as spin wave channels will be implicitly referring to this type of domain wall, unless it is indicated otherwise (as in \lq180 degrees Bloch wall'). To visualize the spin wave dynamics here we demonstrate the results of excitation only of the central core region of the square in a single Landau ground state. This helps to avoid other dynamic effects such as domain wall oscillations and radiation of spin waves from the edges. However, it can be shown that such limitation does not undermine the effect and, even when the whole element is exposed to uniform magnetic field, the same processes can be easily observed. Figure 3 shows examples of magnetization configuration at different times after exposing the square elements to a continuous periodic field $h_{ac}$. It can be clearly seen that as the time progresses the phase of the oscillation  becomes asymmetrically distorted forming a spiral wavefront. Compared to the circular elements however, where the formed spiral is radially uniform, in squares the shape of the front is significantly distorted. As well as the geometric constraints of the shape, the wave propagation is affected by the presence of the domain walls. In both thin and thick elements the walls primarily lead to distortion in local demagnetizing field and thus modifying the dispersion properties. For thicker elements this is a more profound effect, because above a certain limit (approx. 50 nm) the transverse profile of the wall changes significantly, leading to a Bloch type structures with a non-zero out-of plane component. A more (significant/interesting) effect however comes from the dynamics of the walls themselves. Even though the walls are absolutely stationary at this frequency range, due to their confined nature they can host a range of transverse oscillations for the perpendicular component of magnetization alongside the wall. This can lead to a formation of propagating spin-wave modes along the domain walls, or so-called Winter’s modes \cite{PhysRev.124.452}. Our simulations show, that the intensity of these modes can be significantly higher than any other dynamic modes, and thus significantly affect the formation of the spiral waves. Previously it was shown that these modes can arise from the topological inhomogeneities in the sample (e.g. corners or edges)\cite{PhysRevB.84.144406}. In our case, this is very well seen that they originate directly from the core region and propagate with a shorter wavelength, almost twice smaller (see Fig. 3(c)) than the spiral spin waves radiated into the triangular domains of the Landau pattern. The confined spin waves - Winter's magnons - also show opposite phase propagation along the different sides across the core. This later factor helps to couple the spiral waves in the triangular domains together with the magnons in the domain walls. However the difference in the wave numbers leads to a different propagation velocity and consequently the distortion in the circular spiral wavefront that become reminiscent of the geometric shape structure. It can be shown that this works not only for squares, but for shapes of any number of corners, as long as the corners are sharp enough to produce a reasonably confined domain wall. Figures 3 (b) and (c) show how the effect is changing when reducing the angle of the corner. It is interesting to note, that one diagonal seems to be better excited than the other. Numerical results show that in the frequency range where only Winter's magnons are excited, both diagonals show spin waves of similar amplitude. Therefore, the enhancement (attenuation) along one diagonal may be related to constructive (destructive) interference with the corresponding spiral spin waves along the given diagonals or any other standing pattern in the triangular domains, as seen in Fig. 8(b) (both insets). 

\par In order to clarify these findings, we used the analytical dispersion relation from \cite{PhysRevLett.114.247206} for a Winter's magnon propagating in a Bloch wall. In this case, any intrinsic anisotropy is considered to be negligible. We also consider pinned boundary conditions at the edges and the core and assumed an effective internal field value of $H_{eff} = 0$ A/m at the region surrounding the core. Fig. 4 shows a comparison between the dispersion relation along one of the diagonals of the square and that of a Winter's magnon propagating along a 180 (degrees) Bloch wall, defined only by the interplay between exchange and anisotropy. Excitation is applied parallel to one of the edges of the square. As it can be seen from the dispersion, there is a significant mismatch at low ${k}$ values between the analytical solution (solid green line) for a 180 degrees Bloch wall and numerical results from our Landau pattern. We believe this could be due to the fact that the domain wall in the Landau pattern is not entirely a Bloch type wall. Based on the magnetization profile, the region closer to the core shows a reduced out-of-plane component and therefore, closer to a Neel wall configuration (see inset of Fig. 4). In fact, if we consider the propagation of waves in the Neel walls from \cite{PhysRevLett.114.247206}, it can be seen that its dispersion relation is growing faster than that of a Bloch wall. As a result, there is a certain mixing of two regimes, with Neel wall dispersion dominating at low ${k}$ and Bloch wall dispersion at higher ${k}$. A non-zero dipolar field transversal to the square's domain walls, due to the 90 degrees turning as it is suggested in \cite{Lara2017InformationPI}, could also contribute to the mismatches at very low $k$ values. 

\par We continue with the analysis of the effects of rounded corners. This shape is closer in geometry to the real sample, assuming imperfections at the sharp corners happened during fabrication. In such a case, the wavefront becomes modified when approaching the corners (see Fig. 3). The exchange energy density is reduced and the shape anisotropy becomes dominant, resulting in a confined wave with similar wavelength to that of the spiral radiated into the triangular domains. This shows an effective spatial down-chirping of the Winter's magnons wavelength caused by the geometry of the corners. From Fig.1 and Fig.4 we see that the wavelength of Winter's modes is smaller than that of the spiral spin wave (${k_{wm}} > {k_{d}} $), up to a certain frequency. At a frequency close to 19 GHz, it can be deduced that the wavefront becomes a circular spiral, since both wave vectors match (${k_{wm}} = {k_{d}} \approx $ 0.1 rad/nm). This implies that the spiral spin wave can also be modified by just applying oscillating fields of higher frequencies. We ran micromagnetic simulations at 20 GHz CW excitation (not shown here) confirming the latter. Previous works have also suggested to apply biasing fields to modify further the width of the domain walls or \lq exchange channels' for the Winter's magnons \cite{Lara2017InformationPI}.

In the next section, we provide a simple physical explanation of the arising of these characteristic waves in thicker samples.

\subsection{Balance between Dipolar and Exchange energies}
\par In finite magnetic elements in absence of biasing fields, the domain wall formation is mainly governed by their shape and magnetic anisotropy, leading to specific magnetic configuration. Depending on the thickness, either Bloch or Neel walls are preferable as a result of a competition between magnetostatic and exchange energy. This can be intuitively understood as follows: In thin films, out-of-plane rotation of magnetic moments between domains is energetically expensive because of high demagnetizing fields. When the thickness is increased the demagnetizing field is reduced initially at the points of higher magnetic gradients such as boundaries and magnetic singularities, leading to increased out-of-plane components. The same happens with a Neel wall, which at certain thickness (typically around 50 nm for permalloy) starts transferring into a Bloch wall. 
\par Now let us consider an in-plane excitation applied to circular disks. To get a better view on the magnetostatic relations behind these processes, we record the magnitude of magnetization, dipolar and exchange fields across the diameter of the disk. Here we take two case scenarios for a Permalloy nano-disk with thicknesses of 20 nm and 80 nm, where the vortex configuration is reached after relaxation in both situations. Our simulations show a more pronounced out of plane turning in the thicker disks. Interestingly, the orientation of magnetization vectors in thin and thick disks in the ground state resembles the distribution for Neel and Bloch walls, respectively. These differences on the initial distribution contribute to a better excitation of spin waves at the vicinity of the core in thicker disks. From Landau-Lifshitz-Gilbert (LLG) equation \cite{LANDAU199251}, 
\begin{gather}
\frac{d{\mathbf{M}(t)}}{dt} =-\gamma (\mathbf{M}\times \mathbf{H_{eff} })-\lambda\mathbf{M}\times(\mathbf{M}\times \mathbf{H_{eff}}) ,
\end{gather}
where $\gamma$ is the electron gyromagnetic ratio and $\lambda$ a phenomenological damping term ($\lambda$ = $\alpha \gamma / {M_S} $), applying an in-plane excitation to magnetic moments in a thick disk near the core, will enhance the exerted torque. Therefore, spin waves from the vortex core region are not that much excited in thinner discs. These different configurations can also explain why the vortex core diameter becomes larger in thicker samples. In thick disks, since the out of plane rotation of magnetization happens more gradually than in thin samples, the vortex core becomes wider.

\begin{figure}[ht]
\centering 
\includegraphics[trim=0cm 0cm 0cm 0cm, clip=true, width=8cm]{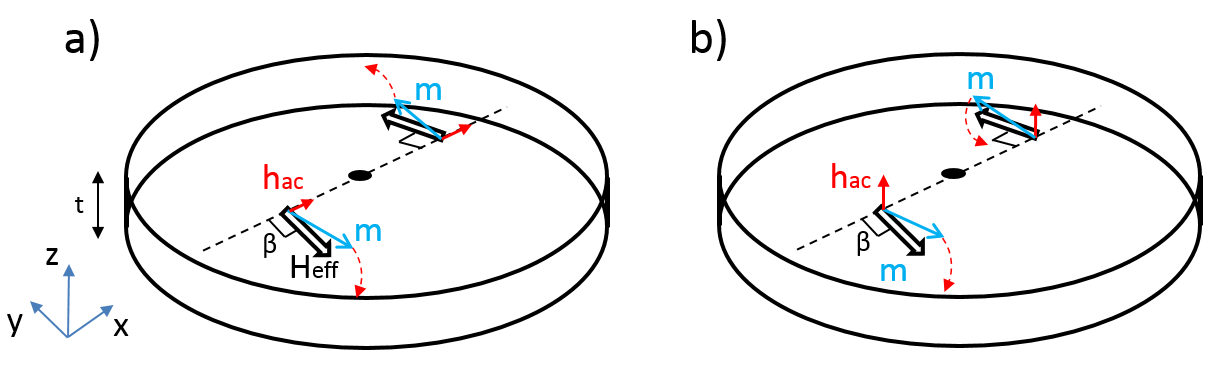}
\caption{Schematic of a thick disk ({t} = 80 nm) with in-plane (a) and out of plane (b) excitation showing counterclockwise precession of magnetic moments (blue arrows) around the effective field (white arrows) at opposite positions referenced to the vortex core (black dot). Dotted red arrow shows the sense of  rotation due to the exerted torque according to Eq. 3, where the out of plane component of magnetization can be deduced to be out of phase by $\pi$ (in-phase) when an in-plane (out of plane) excitation is applied. Vectors are magnified and not scaled for sake of clarity.}  \label{Fields}
\end{figure}

\par The characteristic wavefront of the spiral spin wave can also be easily understood from LLG equation, which dictates the counter-clockwise sense of Larmor precession of magnetization $M$ around $H_{eff}$ (see Fig.5). For  two  vectors in the vicinity of the core, the excitation of applied in-plane  field will lead to out-of-phase precessions on opposite sides of the core, thus leading to the spiral nature of the phase. In case of the out-of-plane excitation, the clock-wise rotation leads to in-phase precession of both vectors, which is then resulted in a radial wave propagation. The asymmetric phase of precession can also explain the existence of the gyrating \lq dip', reported in other studies on thinner films \cite{Kammerer2011MagneticVC}.

\begin{figure}[ht]
\centering 
\includegraphics[trim=0cm 0cm 0cm 0cm, clip=true, width=8cm]{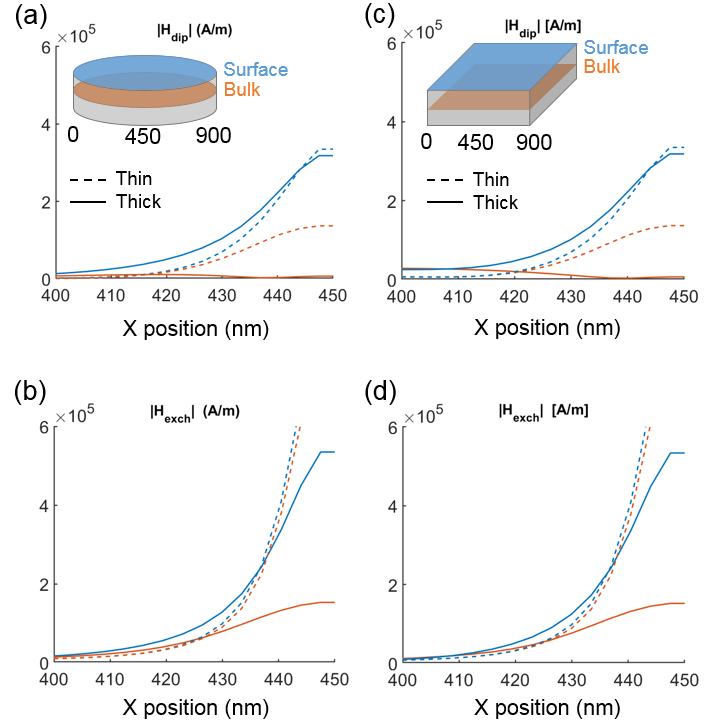}
\caption{Module of magnetostatic and exchange fields in {x} direction across the center for circular disks ((a) and (b)) and square elements ((c) and (d)) of 900 nm diameter and thicknesses 80 nm (solid lines) and 20 nm (dashed lines). Insets show a reference for the sectional cuts for the top and middle layer in thick and thin disks. The x axis direction is chosen to be radial in the circular disks and perpendicular to one edge in the square dot.}  \label{Fields}
\end{figure}

\par Fig. 6 shows the module of dipolar and exchange field at the points surrounding the vortex core in its equilibrium position (from numerical results not shown here, we define that region as $|x-x_{vc}| < 50$ nm, with $x_{vc} = 450$ nm) for thick and thin circular disks ((a) and (b), respectively) and for thick and thin square dots ((c) and (d)). Due to circular (90 degrees) symmetry in the disks (square dots) we only consider the $x$-direction (perpendicular to the dot edge, circular or square) for the following analysis. Firstly, results are similar if not identical for both shapes, so we can infer that similar dynamical behaviour will be found in both element shapes along that direction. From Fig. 6 we can see that, in contrast to the thin disks, dipolar field is drastically reduced in the bulk of the thick disks. From these results, we can infer the following: 1. A weaker demagnetizing field allows magnetic moments to precess out-of-plane with more freedom than when being closer to the surfaces (blue line). This also explains why the spiral spin wave is more intense in the bulk of the thick disk, due to the effective non influence of surface anisotropy and pinning; 2. Dipolar field in the bulk of the thick disk is near to zero. This implies that dipolar interaction with the surfaces is minimal, while exchange interaction in the core region is dominant allowing the exchange-dominated spiral spin waves to occur; 3. Away from the core region and closer to the surfaces, values of the exchange field for thin and thick disks are similar and are very low in contrast to the core region. Via exchange interactions, the spiral spin wave is originated in the middle layers and gets coupled more easily to the neighboring layers. This, also supports the uniformity in phase of the spiral wave across the thickness, which is observed in simulations. In summary, the significant weakening of the demagnetizing field in thicker disks allows the core vicinity in the bulk of the disk to act as a better source for exchange dominated spin waves.

\section{VNA-FMR measurements}

\par In the previous section, we showed that the spiral spin wave emanating from the core is more intense for thicker nano-dots, always assuming  a single vortex state configuration as starting point. Hence, in thick samples, the footprint of these propagating modes is more susceptible of being measured through standard RF techniques.

\subsection{Circular disks}

\begin{figure}[ht]
\centering 
\includegraphics[trim=0cm 0cm 0cm 0cm, clip=true, width=9cm]{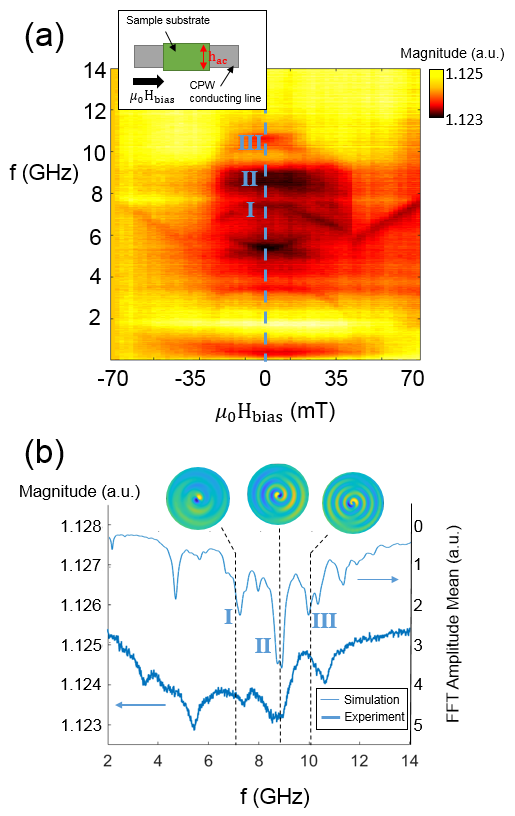}
\caption{VNA-FMR results for circular disks (a) of thickness 80 nm. The FMR main mode can be spotted as well as other low field resonances related with the simulated spiral spin waves (insets in (b)). Cut at zero applied field (dark blue line) compared with the simulated spin wave absorption spectra of a centered vortex (light blue line) (b)}  \label{Fields}
\end{figure}

\par Figure 7(a) shows the experimental absorption spectra and the characteristic absorption peaks from the contribution of the real part of microwave permeability in 80 nm thick dots of 900 nm diameter. The main absorption peaks are present at the frequencies where spiral spin waves can be efficiently excited (these modes are numbered in Fig. 7 as I, II and III), which are in good agreement with our simulated spectra. There are some slight discrepancies, of 4$\%$ at 7 GHz and 5$\%$ at 10 GHz, which can be due to imperfections at the edges which normally lead to a smaller effective sample size. Qualitatively, we observe that some resonances seem to remain stable in a wide range of fields (up to 25 - 30 mT) and therefore, not much influenced by biasing field or dipolar-dipolar interactions between disks and thus ensures that we are observing characteristic modes from the single dots. This follows up from numerical results, where the spiral wave exists as long as the vortex core is not self-annihilated when reaching the edge of the element. In Fig.7(b), the amplitude of the main peak at 8.7 GHz (mode II) corresponds to a standing wave condition in the disk: From numerical simulation (see Fig. 1(b)), we can see that the wavelength of the spiral spin wave at that frequency is 125.6 nm. Considering one half of the disk as a resonant Fabry-Perot cavity, the resonant length ($L$) condition for a set of standing modes of order $n$ would be satisfied given: $L = (1/2)n {\lambda_c}$ where ${\lambda_c}$ is the spin wave wavelength. From Fig. 3(c), for the given wavelength at a frequency of 8.7 GHz and the length of the cavity, the number of anti-nodes found is $n = 7$ which gives a resonant length of $L =$ 440 nm. This is in good agreement with the radius of the disk ($r = d/2 = $450 nm).  It is worth to note that a more realistic 'cavity length' will be shorter due to the width of the vortex core, therefore shorter than the radius of the disk. Similarly, if we repeat for frequencies of 7 GHz and 10 GHz ($n = 5$, mode I; and $n = 9$, mode III respectively), and assuming a resonant cavity length of $L =$ 450 nm, we obtain the same $k$-vectors (or equivalent wavelengths) as in the lower branch of the dispersion diagram from Fig. 1(b), 0.034 rad/nm and 0.069 rad/nm respectively. We also note the consecutive odd values of the measured modes. These are the modes susceptible for activation since the even modes may not be seen due to the symmetry of the signal in the CPW. Therefore, we can confidently correlate the measured modes above 6 GHz to the spiral spin waves emitted from the core vicinity.

\subsection{Squares}
\par Analogously to Sec. IV.A, Fig. 8 shows the experimental results for squares and a comparison with numerical results. From the measured absorption spectra, we can also observe a large resonance peak at 8.8 GHz for a square of dimensions (${L}$) 900 nm by (${t}$) 80 nm, which it is also in very good agreement with our simulations.  Its frequency position falls almost exactly into that of a circular disk of equivalent dimensions where, according to numerical results (see Fig. 2), we are expecting propagation of an intense spiral spin wave.  When compared to the resonance frequency of a perfect square, it is worth to note that there will be an expected frequency mismatch mainly because of imperfections made during the fabrication process, which include: more rounded corners and a longer side length than expected. All these factors may contribute to a frequency down shift which could explain the mismatch and a closer matching to the resonance frequency from the circular disk of 900 nm diameter. 

\begin{figure}[ht]
\centering 
\includegraphics[trim=0cm 0cm 0cm 0cm, clip=true, width=9cm]{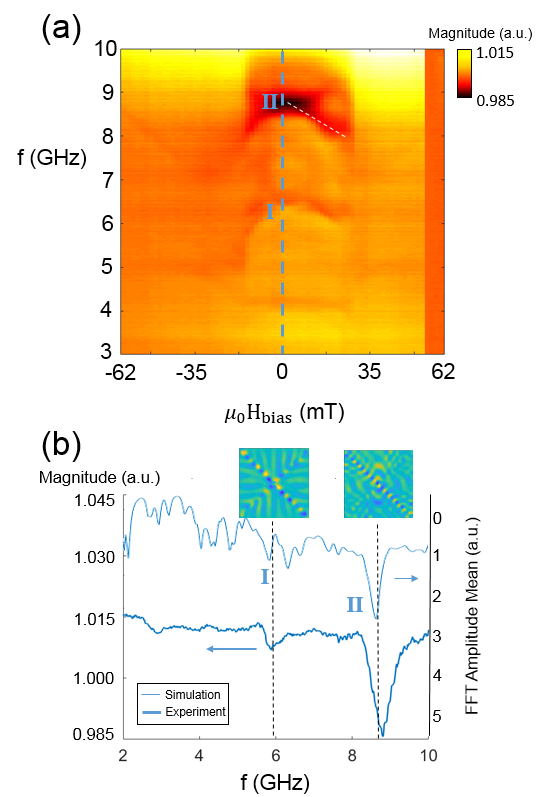}
\caption{VNA-FMR results for square elements (a) of thickness 80 nm. The FMR main mode can be spotted as well as other low field resonances related with the simulated spiral spin waves and Winter's magnons (insets in (b)). (a) shows the splitting of the main resonance peak and white dashed line shows the analytical slope for the downshifted peak. Cut at zero applied field (dark blue line) is compared with the simulated spin wave absorption spectra of a centered vortex (light blue line) (b)}  \label{Fields}
\end{figure}

\par It is interesting that, the resonance peak here is stronger than those found in circular arrays. As previously discussed in section III, this could be related to the formation of 90 degrees domain walls in the sample, characteristic of a Landau pattern, and thus the stronger confinement of the spiral spin waves propagating in them. Considering the spin wave's wavelength from the dispersion diagram in Fig.4 (${\lambda_d} =$ 88.8 nm) and check the resonance condition in a domain wall of length ${l_{dw}}=\sqrt[]{2}L/2 =$ 636 nm, we will obtain, for $n = 14$, a resonant length of ${L} =$ 622 nm which well agrees with the length of half the diagonal of the square. As it happened in the circular disks, a more realistic domain wall length will be always less than half the diagonal of the square. If we repeat the process for the next largest peak, at 6 GHz ($n = 11$ and $n = 8$, respectively, see Fig.8(b)), we obtain an optimal resonant lengths of ${L} =$ 602 nm, which is also in good agreement with half the length of the diagonal, with relative errors of 5$\%$ (0.38$\lambda$). Again, considering a shorter effective length of the domain wall due to the width of the vortex core, the error is even smaller (0.18$\lambda$). Considering all this, the largest measured resonance can be associated with the Winter's propagating modes coherently emitted with the spiral spin wave. The lowest peaks are related with only lower resonances of the Winter's magnons in the domain walls, but not necessarily matched with the spiral spin wave from the core, especially at the lowest frequencies, where the spiral from the core is not excited. 

\par Other reports works have shown previously the splitting of azimuthal spin waves in circular shapes when the vortex core is displaced \cite{PhysRevB.79.174433, PhysRevB.73.134426}. By applying a biasing field parallel to one side of the square, we can shorten or increase the length of opposite domain walls, which splits the resonance into a lower and upper frequency due to the broken symmetry. If the domain walls are of the same length (this is, at $B_{bias} = 0$ mT), the resonance peaks add up and can also match with the absorption peak related to the spiral spin wave radiated into the domains. This enhances the total absorption at that specific frequency (see Fig. 8). We show experimental results of this phenomena for the largest sample (${L} = 900$ nm). For this sample, we obtain an experimental average value of the ratio $\Delta f/ \Delta B=-0.049$ GHz/mT for the peak that shifts down (see Fig. 8(a)). To confirm this analytically, we make the following approximation: If the biasing field is small, i.e. $B_{bias}<<B_{sat}$, we can consider the variation of domain wall length approximately equal to the vortex core displacement ($\Delta s$), $\Delta l_{dw} \approx \Delta s$. We can also put $\Delta f/ \Delta B$ as $(\Delta f/ \Delta s)(\Delta s/ \Delta B)$. From numerical results on thick square disks, we obtained a vortex displacement to biasing field ratio of $\Delta s/ \Delta B=8.59$ nm/mT. Then,we can obtain the relation $\Delta f /\Delta s \approx\Delta f/ \Delta l_{dw}$ from the dispersion relation of discretized spin waves in a normally magnetized film:
\par 
\begin{gather}
f_{n} =f_0+f_M\lambda_{ex} (\frac{n\pi}{l_{dw}})^2
\end{gather}
From numerical results, we obtain a vortex core half-width of 50 nm, we can calculate an effective domain wall length (${l_{dw}}' = (L/\sqrt[]{2}) - 50 = 586$ nm), $n=13$ for a driving frequency of $f=8.7$ GHz (see fig.5b) we derive Eq. 4 with respect to $l_{dw}$ and substitute the corresponding values, obtaining: $\Delta f/ \Delta l_{dw}=-0.0052$ GHz/nm. Finally, and under all the previous approaches, we obtain an analytical value for $\Delta f/ \Delta B =-0.045$ GHz/mT, which is in good agreement with the measured value, with a tolerable error of $9 \% $. While the agreement at low external fields is very good, mismatches at larger biasing fields can be explained since the assumption $\Delta l_{dw} \approx \Delta s$ may no longer hold. By considering this similar variation in the eigenfrequencies for this mode, both in the experiment and the analytical approximation,  we can confidently say that the frequency down-shift due to applying larger biasing fields is a consequence of the standing wave condition of Winter's magnons in the domain walls of square disks. 
\par Figure 9 shows more results on arrays of square elements with smaller length ({L}) at a fixed thickness ({t} = 80 nm): {L} =  600 nm and 500 nm. It was confirmed through electron microscopy (SEM) that the samples had the same separation length, similar to the side length, to avoid dipolar coupling and magnetic interferences among them. It was found that, although the resonance peak shifted to higher frequencies, we also observed that the down shift dependence on the biasing field of the main peak follows as well a similar if not identical slope to the one analytically found for the largest square. From numerical results, the number of nodes found along half the diagonal of the 600 (500) nm square is 9 (8). Given the wavelength of the Winter's magnon excited at frequencies 9.1 (9.4) GHz (see Fig. 4), the optimal resonant length of the domain wall was 399 (338) nm, which was again in good agreement with the ideal length of half the diagonal of the square, 424.3 (353) nm. This gives a tolerable error of 6$\%$ (4$\%$), i.e., an error of 0.28$\lambda$ (0.18$\lambda$), which is even smaller if we consider a more realistic shorter domain wall length due to the width of the vortex core region.
\par For smaller sizes such as the 500 nm and 600 nm squares, reduction of side length in steps of 100 nm (i.e. a reduction of half the diagonal of 70.7 nm), is almost equivalent to eliminating two nodes, i.e. one wavelength (88.8 nm), therefore it seems reasonable that the numbers of the modes are close and consecutive. Assuming that imperfections during fabrication happened regularly for every array and taking all the above into account, we can confidently say that the measured resonances, in thick samples, are associated with the spiral wave emanating from the core vicinity in coherence with Winter's magnons propagating in domain walls.

\begin{figure}[ht]
\centering 
\includegraphics[trim=0cm 0cm 0cm 0cm, clip=true, width=9cm]{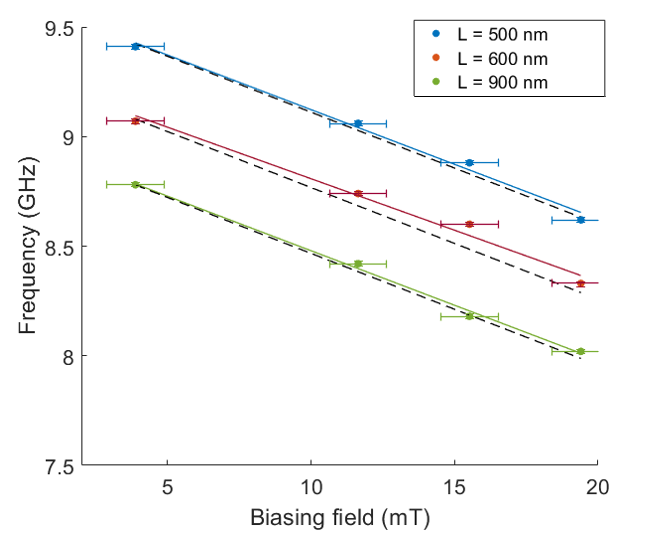}
\caption{VNA-FMR results for squared thick disks of side lengths 900 nm (a), 600 nm (b) and 500 nm (c) applying different biasing fields. Solid lines show a linear fit to experimental data and the error bars are the steps (20 G = 2 mT) of the sweeping of the biasing field. Dashed line is the analytical slope for the downshifted peak.}  \label{Fields}
\end{figure}

\section{Summary}
\par As a consequence of the thickness increase in magnetic thin film elements, the competition between minimization of exchange and dipolar energies dictates the different dynamics and the configuration of the internal magnetic field in the sample. Similarly to the formation of different types of domain walls, we showed the rising of different spin wave modes near the vortex core inhomogeneity and gave an explanation from the perspective of a balance between magnetic energies and the distribution of magnetization vectors. 
\par In the bulk of a thick sample, intensity of dipolar fields is reduced and exchange  fields are enhanced in the vicinity of the vortex core. This creates new magnetization inhomogeneities which causes an outward propagating wave from the vortex core region. If the excitation is applied in-plane, the spin wave shows an out of phase wavefront on opposite sides of the core and hence, conforming a spiral wavefront. In thin disks, the arrangement of magnetic moments considerably reduces the intensity of the generated spin waves. When the pumping field is applied out of plane, a circular wavefront is excited from the core vicinity. In this report, all these scenarios have been explained on the basis of relations used in Landau-Lifshitz equation of motion. Dispersion relations give more insight of the landscape of the excitable modes in circular disks.
\par In conclusion, we showed the possibility of enhancing the amplitude of spiral spin waves in thick nano-patches, improving their detection by means of standard RF techniques such as VNA-FMR. Through this technique, we also show experimental demonstration of absorption peaks related to Winter's magnons traveling along the diagonals of square elements and how the absorption can be enhanced by also exciting simultaneously the spiral spin wave from the core.
\par We hope these results help to better understand the sources of spiral spin waves in confined structures and how to control their dynamical properties. The core vicinity can act as a quasi-punctual source of spiral spin waves, among other shapes of wavefronts, offering a new variety of potential applications for magnetic nanostructures.  
\par The work in Exeter was supported by EPRSC and the CDT in Metamaterials, University of Exeter. The work in Madrid was supported in part by Spanish MINECO (MAT2015-66000-P, RTI2018-095303-B-C55-, EUIN2017-87474, MDM-2014-0377) and Comunidad de Madrid (NANOMAGCOST-CMP2018/NMT-4321). All data created during this research are openly available from the University of Exeter's institutional repository at https://ore.exeter.ac.uk/repository/

\bibliography{library}

\begin{thebibliography}{41}%
\makeatletter
\providecommand \@ifxundefined [1]{%
 \@ifx{#1\undefined}
}%
\providecommand \@ifnum [1]{%
 \ifnum #1\expandafter \@firstoftwo
 \else \expandafter \@secondoftwo
 \fi
}%
\providecommand \@ifx [1]{%
 \ifx #1\expandafter \@firstoftwo
 \else \expandafter \@secondoftwo
 \fi
}%
\providecommand \natexlab [1]{#1}%
\providecommand \enquote  [1]{``#1''}%
\providecommand \bibnamefont  [1]{#1}%
\providecommand \bibfnamefont [1]{#1}%
\providecommand \citenamefont [1]{#1}%
\providecommand \href@noop [0]{\@secondoftwo}%
\providecommand \href [0]{\begingroup \@sanitize@url \@href}%
\providecommand \@href[1]{\@@startlink{#1}\@@href}%
\providecommand \@@href[1]{\endgroup#1\@@endlink}%
\providecommand \@sanitize@url [0]{\catcode `\\12\catcode `\$12\catcode
  `\&12\catcode `\#12\catcode `\^12\catcode `\_12\catcode `\%12\relax}%
\providecommand \@@startlink[1]{}%
\providecommand \@@endlink[0]{}%
\providecommand \url  [0]{\begingroup\@sanitize@url \@url }%
\providecommand \@url [1]{\endgroup\@href {#1}{\urlprefix }}%
\providecommand \urlprefix  [0]{URL }%
\providecommand \Eprint [0]{\href }%
\providecommand \doibase [0]{http://dx.doi.org/}%
\providecommand \selectlanguage [0]{\@gobble}%
\providecommand \bibinfo  [0]{\@secondoftwo}%
\providecommand \bibfield  [0]{\@secondoftwo}%
\providecommand \translation [1]{[#1]}%
\providecommand \BibitemOpen [0]{}%
\providecommand \bibitemStop [0]{}%
\providecommand \bibitemNoStop [0]{.\EOS\space}%
\providecommand \EOS [0]{\spacefactor3000\relax}%
\providecommand \BibitemShut  [1]{\csname bibitem#1\endcsname}%
\let\auto@bib@innerbib\@empty
\bibitem [{\citenamefont {Kruglyak}\ \emph {et~al.}(2010)\citenamefont
  {Kruglyak}, \citenamefont {Demokritov},\ and\ \citenamefont
  {Grundler}}]{0022-3727-43-26-264001}%
  \BibitemOpen
  \bibfield  {author} {\bibinfo {author} {\bibfnamefont {V.~V.}\ \bibnamefont
  {Kruglyak}}, \bibinfo {author} {\bibfnamefont {S.~O.}\ \bibnamefont
  {Demokritov}}, \ and\ \bibinfo {author} {\bibfnamefont {D.}~\bibnamefont
  {Grundler}},\ }\href {http://stacks.iop.org/0022-3727/43/i=26/a=264001}
  {\bibfield  {journal} {\bibinfo  {journal} {Journal of Physics D: Applied
  Physics}\ }\textbf {\bibinfo {volume} {43}},\ \bibinfo {pages} {264001}
  (\bibinfo {year} {2010})}\BibitemShut {NoStop}%
\bibitem [{\citenamefont {Karenowska}\ \emph {et~al.}(2016)\citenamefont
  {Karenowska}, \citenamefont {Chumak}, \citenamefont {Serga},\ and\
  \citenamefont {Hillebrands}}]{Karenowska2016}%
  \BibitemOpen
  \bibfield  {author} {\bibinfo {author} {\bibfnamefont {A.~D.}\ \bibnamefont
  {Karenowska}}, \bibinfo {author} {\bibfnamefont {A.~V.}\ \bibnamefont
  {Chumak}}, \bibinfo {author} {\bibfnamefont {A.~A.}\ \bibnamefont {Serga}}, \
  and\ \bibinfo {author} {\bibfnamefont {B.}~\bibnamefont {Hillebrands}},\
  }\enquote {\bibinfo {title} {Magnon spintronics},}\ in\ \href {\doibase
  10.1007/978-94-007-6892-5_53} {\emph {\bibinfo {booktitle} {Handbook of
  Spintronics}}},\ \bibinfo {editor} {edited by\ \bibinfo {editor}
  {\bibfnamefont {Y.}~\bibnamefont {Xu}}, \bibinfo {editor} {\bibfnamefont
  {D.~D.}\ \bibnamefont {Awschalom}}, \ and\ \bibinfo {editor} {\bibfnamefont
  {J.}~\bibnamefont {Nitta}}}\ (\bibinfo  {publisher} {Springer Netherlands},\
  \bibinfo {address} {Dordrecht},\ \bibinfo {year} {2016})\ pp.\ \bibinfo
  {pages} {1505--1549}\BibitemShut {NoStop}%
\bibitem [{\citenamefont {Hoffmann}\ and\ \citenamefont
  {Bader}(2015)}]{PhysRevApplied.4.047001}%
  \BibitemOpen
  \bibfield  {author} {\bibinfo {author} {\bibfnamefont {A.}~\bibnamefont
  {Hoffmann}}\ and\ \bibinfo {author} {\bibfnamefont {S.~D.}\ \bibnamefont
  {Bader}},\ }\href {\doibase 10.1103/PhysRevApplied.4.047001} {\bibfield
  {journal} {\bibinfo  {journal} {Phys. Rev. Applied}\ }\textbf {\bibinfo
  {volume} {4}},\ \bibinfo {pages} {047001} (\bibinfo {year}
  {2015})}\BibitemShut {NoStop}%
\bibitem [{\citenamefont {Krawczyk}\ and\ \citenamefont
  {Grundler}(2014)}]{0953-8984-26-12-123202}%
  \BibitemOpen
  \bibfield  {author} {\bibinfo {author} {\bibfnamefont {M.}~\bibnamefont
  {Krawczyk}}\ and\ \bibinfo {author} {\bibfnamefont {D.}~\bibnamefont
  {Grundler}},\ }\href {http://stacks.iop.org/0953-8984/26/i=12/a=123202}
  {\bibfield  {journal} {\bibinfo  {journal} {Journal of Physics: Condensed
  Matter}\ }\textbf {\bibinfo {volume} {26}},\ \bibinfo {pages} {123202}
  (\bibinfo {year} {2014})}\BibitemShut {NoStop}%
\bibitem [{\citenamefont {Whitehead}\ \emph {et~al.}(2017)\citenamefont
  {Whitehead}, \citenamefont {Horsley}, \citenamefont {Philbin}, \citenamefont
  {Kuchko},\ and\ \citenamefont {Kruglyak}}]{PhysRevB.96.064415}%
  \BibitemOpen
  \bibfield  {author} {\bibinfo {author} {\bibfnamefont {N.~J.}\ \bibnamefont
  {Whitehead}}, \bibinfo {author} {\bibfnamefont {S.~A.~R.}\ \bibnamefont
  {Horsley}}, \bibinfo {author} {\bibfnamefont {T.~G.}\ \bibnamefont
  {Philbin}}, \bibinfo {author} {\bibfnamefont {A.~N.}\ \bibnamefont {Kuchko}},
  \ and\ \bibinfo {author} {\bibfnamefont {V.~V.}\ \bibnamefont {Kruglyak}},\
  }\href {\doibase 10.1103/PhysRevB.96.064415} {\bibfield  {journal} {\bibinfo
  {journal} {Phys. Rev. B}\ }\textbf {\bibinfo {volume} {96}},\ \bibinfo
  {pages} {064415} (\bibinfo {year} {2017})}\BibitemShut {NoStop}%
\bibitem [{\citenamefont {Davies}\ \emph {et~al.}(2015)\citenamefont {Davies},
  \citenamefont {Francis}, \citenamefont {Sadovnikov}, \citenamefont
  {Chertopalov}, \citenamefont {Bryan}, \citenamefont {Grishin}, \citenamefont
  {Allwood}, \citenamefont {Sharaevskii}, \citenamefont {Nikitov},\ and\
  \citenamefont {Kruglyak}}]{article}%
  \BibitemOpen
  \bibfield  {author} {\bibinfo {author} {\bibfnamefont {C.}~\bibnamefont
  {Davies}}, \bibinfo {author} {\bibfnamefont {A.}~\bibnamefont {Francis}},
  \bibinfo {author} {\bibfnamefont {A.}~\bibnamefont {Sadovnikov}}, \bibinfo
  {author} {\bibfnamefont {S.}~\bibnamefont {Chertopalov}}, \bibinfo {author}
  {\bibfnamefont {M.}~\bibnamefont {Bryan}}, \bibinfo {author} {\bibfnamefont
  {S.}~\bibnamefont {Grishin}}, \bibinfo {author} {\bibfnamefont
  {D.}~\bibnamefont {Allwood}}, \bibinfo {author} {\bibfnamefont
  {Y.}~\bibnamefont {Sharaevskii}}, \bibinfo {author} {\bibfnamefont
  {S.}~\bibnamefont {Nikitov}}, \ and\ \bibinfo {author} {\bibfnamefont
  {V.}~\bibnamefont {Kruglyak}},\ }\href@noop {} {\bibfield  {journal}
  {\bibinfo  {journal} {Phy. Rev. B}\ }\textbf {\bibinfo {volume} {92}},\
  \bibinfo {pages} {020408} (\bibinfo {year} {2015})}\BibitemShut {NoStop}%
\bibitem [{\citenamefont {Bailleul}\ \emph {et~al.}(2007)\citenamefont
  {Bailleul}, \citenamefont {H\"ollinger}, \citenamefont {Perzlmaier},\ and\
  \citenamefont {Fermon}}]{PhysRevB.76.224401}%
  \BibitemOpen
  \bibfield  {author} {\bibinfo {author} {\bibfnamefont {M.}~\bibnamefont
  {Bailleul}}, \bibinfo {author} {\bibfnamefont {R.}~\bibnamefont
  {H\"ollinger}}, \bibinfo {author} {\bibfnamefont {K.}~\bibnamefont
  {Perzlmaier}}, \ and\ \bibinfo {author} {\bibfnamefont {C.}~\bibnamefont
  {Fermon}},\ }\href {\doibase 10.1103/PhysRevB.76.224401} {\bibfield
  {journal} {\bibinfo  {journal} {Phys. Rev. B}\ }\textbf {\bibinfo {volume}
  {76}},\ \bibinfo {pages} {224401} (\bibinfo {year} {2007})}\BibitemShut
  {NoStop}%
\bibitem [{\citenamefont {Bailleul}\ \emph {et~al.}(2006)\citenamefont
  {Bailleul}, \citenamefont {H\"ollinger},\ and\ \citenamefont
  {Fermon}}]{PhysRevB.73.104424}%
  \BibitemOpen
  \bibfield  {author} {\bibinfo {author} {\bibfnamefont {M.}~\bibnamefont
  {Bailleul}}, \bibinfo {author} {\bibfnamefont {R.}~\bibnamefont
  {H\"ollinger}}, \ and\ \bibinfo {author} {\bibfnamefont {C.}~\bibnamefont
  {Fermon}},\ }\href {\doibase 10.1103/PhysRevB.73.104424} {\bibfield
  {journal} {\bibinfo  {journal} {Phys. Rev. B}\ }\textbf {\bibinfo {volume}
  {73}},\ \bibinfo {pages} {104424} (\bibinfo {year} {2006})}\BibitemShut
  {NoStop}%
\bibitem [{\citenamefont {Perzlmaier}\ \emph {et~al.}(2005)\citenamefont
  {Perzlmaier}, \citenamefont {Buess}, \citenamefont {Back}, \citenamefont
  {E~Demidov}, \citenamefont {Hillebrands},\ and\ \citenamefont
  {Demokritov}}]{articlesquareseigenmodes}%
  \BibitemOpen
  \bibfield  {author} {\bibinfo {author} {\bibfnamefont {K.}~\bibnamefont
  {Perzlmaier}}, \bibinfo {author} {\bibfnamefont {M.}~\bibnamefont {Buess}},
  \bibinfo {author} {\bibfnamefont {C.}~\bibnamefont {Back}}, \bibinfo {author}
  {\bibfnamefont {V.}~\bibnamefont {E~Demidov}}, \bibinfo {author}
  {\bibfnamefont {B.}~\bibnamefont {Hillebrands}}, \ and\ \bibinfo {author}
  {\bibfnamefont {S.}~\bibnamefont {Demokritov}},\ }\href@noop {} {\bibfield
  {journal} {\bibinfo  {journal} {Phys. Rev. Lett.}\ }\textbf {\bibinfo
  {volume} {94}},\ \bibinfo {pages} {057202} (\bibinfo {year}
  {2005})}\BibitemShut {NoStop}%
\bibitem [{\citenamefont {Schlomann}(1964)}]{doi:10.1063/1.1713058}%
  \BibitemOpen
  \bibfield  {author} {\bibinfo {author} {\bibfnamefont {E.}~\bibnamefont
  {Schlomann}},\ }\href {\doibase 10.1063/1.1713058} {\bibfield  {journal}
  {\bibinfo  {journal} {Journal of Applied Physics}\ }\textbf {\bibinfo
  {volume} {35}},\ \bibinfo {pages} {159} (\bibinfo {year} {1964})}\BibitemShut
  {NoStop}%
\bibitem [{\citenamefont {Davies}\ \emph {et~al.}(2017)\citenamefont {Davies},
  \citenamefont {Poimanov},\ and\ \citenamefont
  {Kruglyak}}]{PhysRevB.96.094430}%
  \BibitemOpen
  \bibfield  {author} {\bibinfo {author} {\bibfnamefont {C.~S.}\ \bibnamefont
  {Davies}}, \bibinfo {author} {\bibfnamefont {V.~D.}\ \bibnamefont
  {Poimanov}}, \ and\ \bibinfo {author} {\bibfnamefont {V.~V.}\ \bibnamefont
  {Kruglyak}},\ }\href {\doibase 10.1103/PhysRevB.96.094430} {\bibfield
  {journal} {\bibinfo  {journal} {Phys. Rev. B}\ }\textbf {\bibinfo {volume}
  {96}},\ \bibinfo {pages} {094430} (\bibinfo {year} {2017})}\BibitemShut
  {NoStop}%
\bibitem [{\citenamefont {Mushenok}\ \emph {et~al.}(2017)\citenamefont
  {Mushenok}, \citenamefont {Dost}, \citenamefont {Davies}, \citenamefont
  {Allwood}, \citenamefont {Inkson}, \citenamefont {Hrkac},\ and\ \citenamefont
  {Kruglyak}}]{doi:10.1063/1.4995991}%
  \BibitemOpen
  \bibfield  {author} {\bibinfo {author} {\bibfnamefont {F.~B.}\ \bibnamefont
  {Mushenok}}, \bibinfo {author} {\bibfnamefont {R.}~\bibnamefont {Dost}},
  \bibinfo {author} {\bibfnamefont {C.~S.}\ \bibnamefont {Davies}}, \bibinfo
  {author} {\bibfnamefont {D.~A.}\ \bibnamefont {Allwood}}, \bibinfo {author}
  {\bibfnamefont {B.~J.}\ \bibnamefont {Inkson}}, \bibinfo {author}
  {\bibfnamefont {G.}~\bibnamefont {Hrkac}}, \ and\ \bibinfo {author}
  {\bibfnamefont {V.~V.}\ \bibnamefont {Kruglyak}},\ }\href {\doibase
  10.1063/1.4995991} {\bibfield  {journal} {\bibinfo  {journal} {Applied
  Physics Letters}\ }\textbf {\bibinfo {volume} {111}},\ \bibinfo {pages}
  {042404} (\bibinfo {year} {2017})}\BibitemShut {NoStop}%
\bibitem [{\citenamefont {Lara}\ \emph {et~al.}(2013)\citenamefont {Lara},
  \citenamefont {Metlushko},\ and\ \citenamefont {Aliev}}]{propedges}%
  \BibitemOpen
  \bibfield  {author} {\bibinfo {author} {\bibfnamefont {A.}~\bibnamefont
  {Lara}}, \bibinfo {author} {\bibfnamefont {V.}~\bibnamefont {Metlushko}}, \
  and\ \bibinfo {author} {\bibfnamefont {F.~G.}\ \bibnamefont {Aliev}},\ }\href
  {\doibase 10.1063/1.4839315} {\bibfield  {journal} {\bibinfo  {journal}
  {Journal of Applied Physics}\ }\textbf {\bibinfo {volume} {114}},\ \bibinfo
  {pages} {213905} (\bibinfo {year} {2013})}\BibitemShut {NoStop}%
\bibitem [{\citenamefont {Kostylev}\ \emph {et~al.}(2005)\citenamefont
  {Kostylev}, \citenamefont {Serga}, \citenamefont {Schneider}, \citenamefont
  {Leven},\ and\ \citenamefont {Hillebrands}}]{doi:10.1063/1.2089147}%
  \BibitemOpen
  \bibfield  {author} {\bibinfo {author} {\bibfnamefont {M.~P.}\ \bibnamefont
  {Kostylev}}, \bibinfo {author} {\bibfnamefont {A.~A.}\ \bibnamefont {Serga}},
  \bibinfo {author} {\bibfnamefont {T.}~\bibnamefont {Schneider}}, \bibinfo
  {author} {\bibfnamefont {B.}~\bibnamefont {Leven}}, \ and\ \bibinfo {author}
  {\bibfnamefont {B.}~\bibnamefont {Hillebrands}},\ }\href {\doibase
  10.1063/1.2089147} {\bibfield  {journal} {\bibinfo  {journal} {Applied
  Physics Letters}\ }\textbf {\bibinfo {volume} {87}},\ \bibinfo {pages}
  {153501} (\bibinfo {year} {2005})}\BibitemShut {NoStop}%
\bibitem [{\citenamefont {Lee}\ and\ \citenamefont
  {Kim}(2008)}]{doi:10.1063/1.2975235}%
  \BibitemOpen
  \bibfield  {author} {\bibinfo {author} {\bibfnamefont {K.-S.}\ \bibnamefont
  {Lee}}\ and\ \bibinfo {author} {\bibfnamefont {S.-K.}\ \bibnamefont {Kim}},\
  }\href {\doibase 10.1063/1.2975235} {\bibfield  {journal} {\bibinfo
  {journal} {Journal of Applied Physics}\ }\textbf {\bibinfo {volume} {104}},\
  \bibinfo {pages} {053909} (\bibinfo {year} {2008})}\BibitemShut {NoStop}%
\bibitem [{\citenamefont {Chumak}\ \emph {et~al.}(2014)\citenamefont {Chumak},
  \citenamefont {Serga},\ and\ \citenamefont
  {Hillebrands}}]{Chumak2014MagnonTF}%
  \BibitemOpen
  \bibfield  {author} {\bibinfo {author} {\bibfnamefont {A.~V.}\ \bibnamefont
  {Chumak}}, \bibinfo {author} {\bibfnamefont {A.~A.}\ \bibnamefont {Serga}}, \
  and\ \bibinfo {author} {\bibfnamefont {B.}~\bibnamefont {Hillebrands}},\
  }\href@noop {} {\bibfield  {journal} {\bibinfo  {journal} {Nature
  communications}\ }\textbf {\bibinfo {volume} {5}},\ \bibinfo {pages} {4700}
  (\bibinfo {year} {2014})}\BibitemShut {NoStop}%
\bibitem [{\citenamefont {Lan}\ \emph {et~al.}(2015)\citenamefont {Lan},
  \citenamefont {Yu}, \citenamefont {Wu},\ and\ \citenamefont
  {Xiao}}]{PhysRevX.5.041049}%
  \BibitemOpen
  \bibfield  {author} {\bibinfo {author} {\bibfnamefont {J.}~\bibnamefont
  {Lan}}, \bibinfo {author} {\bibfnamefont {W.}~\bibnamefont {Yu}}, \bibinfo
  {author} {\bibfnamefont {R.}~\bibnamefont {Wu}}, \ and\ \bibinfo {author}
  {\bibfnamefont {J.}~\bibnamefont {Xiao}},\ }\href {\doibase
  10.1103/PhysRevX.5.041049} {\bibfield  {journal} {\bibinfo  {journal} {Phys.
  Rev. X}\ }\textbf {\bibinfo {volume} {5}},\ \bibinfo {pages} {041049}
  (\bibinfo {year} {2015})}\BibitemShut {NoStop}%
\bibitem [{\citenamefont {Vogt}\ \emph {et~al.}(2012)\citenamefont {Vogt},
  \citenamefont {Schultheiss}, \citenamefont {Jain}, \citenamefont {Pearson},
  \citenamefont {Hoffmann}, \citenamefont {Bader},\ and\ \citenamefont
  {Hillebrands}}]{doi:10.1063/1.4738887}%
  \BibitemOpen
  \bibfield  {author} {\bibinfo {author} {\bibfnamefont {K.}~\bibnamefont
  {Vogt}}, \bibinfo {author} {\bibfnamefont {H.}~\bibnamefont {Schultheiss}},
  \bibinfo {author} {\bibfnamefont {S.}~\bibnamefont {Jain}}, \bibinfo {author}
  {\bibfnamefont {J.~E.}\ \bibnamefont {Pearson}}, \bibinfo {author}
  {\bibfnamefont {A.}~\bibnamefont {Hoffmann}}, \bibinfo {author}
  {\bibfnamefont {S.~D.}\ \bibnamefont {Bader}}, \ and\ \bibinfo {author}
  {\bibfnamefont {B.}~\bibnamefont {Hillebrands}},\ }\href {\doibase
  10.1063/1.4738887} {\bibfield  {journal} {\bibinfo  {journal} {Applied
  Physics Letters}\ }\textbf {\bibinfo {volume} {101}},\ \bibinfo {pages}
  {042410} (\bibinfo {year} {2012})}\BibitemShut {NoStop}%
\bibitem [{\citenamefont {Xing}\ \emph {et~al.}(2013)\citenamefont {Xing},
  \citenamefont {Yu}, \citenamefont {Li},\ and\ \citenamefont
  {Huang}}]{articlebending}%
  \BibitemOpen
  \bibfield  {author} {\bibinfo {author} {\bibfnamefont {X.}~\bibnamefont
  {Xing}}, \bibinfo {author} {\bibfnamefont {Y.}~\bibnamefont {Yu}}, \bibinfo
  {author} {\bibfnamefont {S.}~\bibnamefont {Li}}, \ and\ \bibinfo {author}
  {\bibfnamefont {X.}~\bibnamefont {Huang}},\ }\href {\doibase
  10.1038/srep02958} {\bibfield  {journal} {\bibinfo  {journal} {Scientific
  reports}\ }\textbf {\bibinfo {volume} {3}},\ \bibinfo {pages} {2958}
  (\bibinfo {year} {2013})}\BibitemShut {NoStop}%
\bibitem [{\citenamefont {Wintz}\ \emph {et~al.}(2016)\citenamefont {Wintz},
  \citenamefont {Tiberkevich}, \citenamefont {Weigand}, \citenamefont {Raabe},
  \citenamefont {Lindner}, \citenamefont {Erbe}, \citenamefont {Slavin},\ and\
  \citenamefont {Fassbender}}]{articlewintz}%
  \BibitemOpen
  \bibfield  {author} {\bibinfo {author} {\bibfnamefont {S.}~\bibnamefont
  {Wintz}}, \bibinfo {author} {\bibfnamefont {V.}~\bibnamefont {Tiberkevich}},
  \bibinfo {author} {\bibfnamefont {M.}~\bibnamefont {Weigand}}, \bibinfo
  {author} {\bibfnamefont {J.}~\bibnamefont {Raabe}}, \bibinfo {author}
  {\bibfnamefont {J.}~\bibnamefont {Lindner}}, \bibinfo {author} {\bibfnamefont
  {A.}~\bibnamefont {Erbe}}, \bibinfo {author} {\bibfnamefont {A.}~\bibnamefont
  {Slavin}}, \ and\ \bibinfo {author} {\bibfnamefont {J.}~\bibnamefont
  {Fassbender}},\ }\href@noop {} {\bibfield  {journal} {\bibinfo  {journal}
  {Nature Nanotechnology}\ }\textbf {\bibinfo {volume} {11}},\ \bibinfo {pages}
  {948–953} (\bibinfo {year} {2016})}\BibitemShut {NoStop}%
\bibitem [{\citenamefont {Dieterle}\ \emph {et~al.}(2019)\citenamefont
  {Dieterle}, \citenamefont {F\"orster}, \citenamefont {Stoll}, \citenamefont
  {Semisalova}, \citenamefont {Finizio}, \citenamefont {Gangwar}, \citenamefont
  {Weigand}, \citenamefont {Noske}, \citenamefont {F\"ahnle}, \citenamefont
  {Bykova}, \citenamefont {Gr\"afe}, \citenamefont {Bozhko}, \citenamefont
  {Musiienko-Shmarova}, \citenamefont {Tiberkevich}, \citenamefont {Slavin},
  \citenamefont {Back}, \citenamefont {Raabe}, \citenamefont {Schutz},\ and\
  \citenamefont {Wintz}}]{2017arXiv171200681D}%
  \BibitemOpen
  \bibfield  {author} {\bibinfo {author} {\bibfnamefont {G.}~\bibnamefont
  {Dieterle}}, \bibinfo {author} {\bibfnamefont {J.}~\bibnamefont {F\"orster}},
  \bibinfo {author} {\bibfnamefont {H.}~\bibnamefont {Stoll}}, \bibinfo
  {author} {\bibfnamefont {A.~S.}\ \bibnamefont {Semisalova}}, \bibinfo
  {author} {\bibfnamefont {S.}~\bibnamefont {Finizio}}, \bibinfo {author}
  {\bibfnamefont {A.}~\bibnamefont {Gangwar}}, \bibinfo {author} {\bibfnamefont
  {M.}~\bibnamefont {Weigand}}, \bibinfo {author} {\bibfnamefont
  {M.}~\bibnamefont {Noske}}, \bibinfo {author} {\bibfnamefont
  {M.}~\bibnamefont {F\"ahnle}}, \bibinfo {author} {\bibfnamefont
  {I.}~\bibnamefont {Bykova}}, \bibinfo {author} {\bibfnamefont
  {J.}~\bibnamefont {Gr\"afe}}, \bibinfo {author} {\bibfnamefont {D.~A.}\
  \bibnamefont {Bozhko}}, \bibinfo {author} {\bibfnamefont {H.~Y.}\
  \bibnamefont {Musiienko-Shmarova}}, \bibinfo {author} {\bibfnamefont
  {V.}~\bibnamefont {Tiberkevich}}, \bibinfo {author} {\bibfnamefont {A.~N.}\
  \bibnamefont {Slavin}}, \bibinfo {author} {\bibfnamefont {C.~H.}\
  \bibnamefont {Back}}, \bibinfo {author} {\bibfnamefont {J.}~\bibnamefont
  {Raabe}}, \bibinfo {author} {\bibfnamefont {G.}~\bibnamefont {Schutz}}, \
  and\ \bibinfo {author} {\bibfnamefont {S.}~\bibnamefont {Wintz}},\
  }\href@noop {} {\bibfield  {journal} {\bibinfo  {journal} {Phys. Rev. Lett.}\
  }\textbf {\bibinfo {volume} {122}},\ \bibinfo {pages} {117202} (\bibinfo
  {year} {2019})}\BibitemShut {NoStop}%
\bibitem [{\citenamefont {Kammerer}\ \emph {et~al.}(2011)\citenamefont
  {Kammerer}, \citenamefont {Weigand}, \citenamefont {Curcic}, \citenamefont
  {Noske}, \citenamefont {Sproll}, \citenamefont {Vansteenkiste}, \citenamefont
  {Waeyenberge}, \citenamefont {Stoll}, \citenamefont {Woltersdorf},
  \citenamefont {Back},\ and\ \citenamefont {Schutz}}]{Kammerer2011MagneticVC}%
  \BibitemOpen
  \bibfield  {author} {\bibinfo {author} {\bibfnamefont {M.}~\bibnamefont
  {Kammerer}}, \bibinfo {author} {\bibfnamefont {M.}~\bibnamefont {Weigand}},
  \bibinfo {author} {\bibfnamefont {M.}~\bibnamefont {Curcic}}, \bibinfo
  {author} {\bibfnamefont {M.}~\bibnamefont {Noske}}, \bibinfo {author}
  {\bibfnamefont {M.}~\bibnamefont {Sproll}}, \bibinfo {author} {\bibfnamefont
  {A.}~\bibnamefont {Vansteenkiste}}, \bibinfo {author} {\bibfnamefont {B.~V.}\
  \bibnamefont {Waeyenberge}}, \bibinfo {author} {\bibfnamefont
  {H.}~\bibnamefont {Stoll}}, \bibinfo {author} {\bibfnamefont
  {G.}~\bibnamefont {Woltersdorf}}, \bibinfo {author} {\bibfnamefont {C.~H.}\
  \bibnamefont {Back}}, \ and\ \bibinfo {author} {\bibfnamefont
  {G.}~\bibnamefont {Schutz}},\ }\href@noop {} {\bibfield  {journal} {\bibinfo
  {journal} {Nature communications}\ ,\ \bibinfo {pages} {279}} (\bibinfo
  {year} {2011})}\BibitemShut {NoStop}%
\bibitem [{\citenamefont {Stoll}\ \emph {et~al.}(2015)\citenamefont {Stoll},
  \citenamefont {Noske}, \citenamefont {Weigand}, \citenamefont {Richter},
  \citenamefont {Kruger}, \citenamefont {Reeve}, \citenamefont {Hanze},
  \citenamefont {Adolff}, \citenamefont {Stein}, \citenamefont {Meier},
  \citenamefont {Klaui},\ and\ \citenamefont
  {Schutz}}]{10.3389/fphy.2015.00026}%
  \BibitemOpen
  \bibfield  {author} {\bibinfo {author} {\bibfnamefont {H.}~\bibnamefont
  {Stoll}}, \bibinfo {author} {\bibfnamefont {M.}~\bibnamefont {Noske}},
  \bibinfo {author} {\bibfnamefont {M.}~\bibnamefont {Weigand}}, \bibinfo
  {author} {\bibfnamefont {K.}~\bibnamefont {Richter}}, \bibinfo {author}
  {\bibfnamefont {B.}~\bibnamefont {Kruger}}, \bibinfo {author} {\bibfnamefont
  {R.~M.}\ \bibnamefont {Reeve}}, \bibinfo {author} {\bibfnamefont
  {M.}~\bibnamefont {Hanze}}, \bibinfo {author} {\bibfnamefont {C.~F.}\
  \bibnamefont {Adolff}}, \bibinfo {author} {\bibfnamefont {F.-U.}\
  \bibnamefont {Stein}}, \bibinfo {author} {\bibfnamefont {G.}~\bibnamefont
  {Meier}}, \bibinfo {author} {\bibfnamefont {M.}~\bibnamefont {Klaui}}, \ and\
  \bibinfo {author} {\bibfnamefont {G.}~\bibnamefont {Schutz}},\ }\href
  {\doibase 10.3389/fphy.2015.00026} {\bibfield  {journal} {\bibinfo  {journal}
  {Frontiers in Physics}\ }\textbf {\bibinfo {volume} {3}},\ \bibinfo {pages}
  {26} (\bibinfo {year} {2015})}\BibitemShut {NoStop}%
\bibitem [{\citenamefont {Verba}\ \emph {et~al.}(2016)\citenamefont {Verba},
  \citenamefont {Hierro-Rodriguez}, \citenamefont {Navas}, \citenamefont
  {Ding}, \citenamefont {Liu}, \citenamefont {Adeyeye}, \citenamefont
  {Guslienko},\ and\ \citenamefont {Kakazei}}]{PhysRevB.93.214437}%
  \BibitemOpen
  \bibfield  {author} {\bibinfo {author} {\bibfnamefont {R.~V.}\ \bibnamefont
  {Verba}}, \bibinfo {author} {\bibfnamefont {A.}~\bibnamefont
  {Hierro-Rodriguez}}, \bibinfo {author} {\bibfnamefont {D.}~\bibnamefont
  {Navas}}, \bibinfo {author} {\bibfnamefont {J.}~\bibnamefont {Ding}},
  \bibinfo {author} {\bibfnamefont {X.~M.}\ \bibnamefont {Liu}}, \bibinfo
  {author} {\bibfnamefont {A.~O.}\ \bibnamefont {Adeyeye}}, \bibinfo {author}
  {\bibfnamefont {K.~Y.}\ \bibnamefont {Guslienko}}, \ and\ \bibinfo {author}
  {\bibfnamefont {G.~N.}\ \bibnamefont {Kakazei}},\ }\href {\doibase
  10.1103/PhysRevB.93.214437} {\bibfield  {journal} {\bibinfo  {journal} {Phys.
  Rev. B}\ }\textbf {\bibinfo {volume} {93}},\ \bibinfo {pages} {214437}
  (\bibinfo {year} {2016})}\BibitemShut {NoStop}%
\bibitem [{\citenamefont {Neudecker}\ \emph {et~al.}(2006)\citenamefont
  {Neudecker}, \citenamefont {Perzlmaier}, \citenamefont {Hoffmann},
  \citenamefont {Woltersdorf}, \citenamefont {Buess}, \citenamefont {Weiss},\
  and\ \citenamefont {Back}}]{PhysRevB.73.134426}%
  \BibitemOpen
  \bibfield  {author} {\bibinfo {author} {\bibfnamefont {I.}~\bibnamefont
  {Neudecker}}, \bibinfo {author} {\bibfnamefont {K.}~\bibnamefont
  {Perzlmaier}}, \bibinfo {author} {\bibfnamefont {F.}~\bibnamefont
  {Hoffmann}}, \bibinfo {author} {\bibfnamefont {G.}~\bibnamefont
  {Woltersdorf}}, \bibinfo {author} {\bibfnamefont {M.}~\bibnamefont {Buess}},
  \bibinfo {author} {\bibfnamefont {D.}~\bibnamefont {Weiss}}, \ and\ \bibinfo
  {author} {\bibfnamefont {C.~H.}\ \bibnamefont {Back}},\ }\href {\doibase
  10.1103/PhysRevB.73.134426} {\bibfield  {journal} {\bibinfo  {journal} {Phys.
  Rev. B}\ }\textbf {\bibinfo {volume} {73}},\ \bibinfo {pages} {134426}
  (\bibinfo {year} {2006})}\BibitemShut {NoStop}%
\bibitem [{\citenamefont {Vansteenkiste}\ \emph {et~al.}(2014)\citenamefont
  {Vansteenkiste}, \citenamefont {Leliaert}, \citenamefont {Dvornik},
  \citenamefont {Helsen}, \citenamefont {Garcia-Sanchez},\ and\ \citenamefont
  {Van~Waeyenberge}}]{doi:10.1063/1.4899186}%
  \BibitemOpen
  \bibfield  {author} {\bibinfo {author} {\bibfnamefont {A.}~\bibnamefont
  {Vansteenkiste}}, \bibinfo {author} {\bibfnamefont {J.}~\bibnamefont
  {Leliaert}}, \bibinfo {author} {\bibfnamefont {M.}~\bibnamefont {Dvornik}},
  \bibinfo {author} {\bibfnamefont {M.}~\bibnamefont {Helsen}}, \bibinfo
  {author} {\bibfnamefont {F.}~\bibnamefont {Garcia-Sanchez}}, \ and\ \bibinfo
  {author} {\bibfnamefont {B.}~\bibnamefont {Van~Waeyenberge}},\ }\href
  {\doibase 10.1063/1.4899186} {\bibfield  {journal} {\bibinfo  {journal} {AIP
  Advances}\ }\textbf {\bibinfo {volume} {4}},\ \bibinfo {pages} {107133}
  (\bibinfo {year} {2014})}\BibitemShut {NoStop}%
\bibitem [{\citenamefont {Aliev}\ \emph {et~al.}(2009)\citenamefont {Aliev},
  \citenamefont {Sierra}, \citenamefont {Awad}, \citenamefont {Kakazei},
  \citenamefont {Han}, \citenamefont {Kim}, \citenamefont {Metlushko},
  \citenamefont {Ilic},\ and\ \citenamefont {Guslienko}}]{PhysRevB.79.174433}%
  \BibitemOpen
  \bibfield  {author} {\bibinfo {author} {\bibfnamefont {F.~G.}\ \bibnamefont
  {Aliev}}, \bibinfo {author} {\bibfnamefont {J.~F.}\ \bibnamefont {Sierra}},
  \bibinfo {author} {\bibfnamefont {A.~A.}\ \bibnamefont {Awad}}, \bibinfo
  {author} {\bibfnamefont {G.~N.}\ \bibnamefont {Kakazei}}, \bibinfo {author}
  {\bibfnamefont {D.-S.}\ \bibnamefont {Han}}, \bibinfo {author} {\bibfnamefont
  {S.-K.}\ \bibnamefont {Kim}}, \bibinfo {author} {\bibfnamefont
  {V.}~\bibnamefont {Metlushko}}, \bibinfo {author} {\bibfnamefont
  {B.}~\bibnamefont {Ilic}}, \ and\ \bibinfo {author} {\bibfnamefont {K.~Y.}\
  \bibnamefont {Guslienko}},\ }\href@noop {} {\bibfield  {journal} {\bibinfo
  {journal} {Phys. Rev. B}\ }\textbf {\bibinfo {volume} {79}},\ \bibinfo
  {pages} {174433} (\bibinfo {year} {2009})}\BibitemShut {NoStop}%
\bibitem [{\citenamefont {Taurel}\ \emph {et~al.}(2016)\citenamefont {Taurel},
  \citenamefont {Valet}, \citenamefont {Naletov}, \citenamefont {Vukadinovic},
  \citenamefont {de~Loubens},\ and\ \citenamefont
  {Klein}}]{PhysRevB.93.184427}%
  \BibitemOpen
  \bibfield  {author} {\bibinfo {author} {\bibfnamefont {B.}~\bibnamefont
  {Taurel}}, \bibinfo {author} {\bibfnamefont {T.}~\bibnamefont {Valet}},
  \bibinfo {author} {\bibfnamefont {V.~V.}\ \bibnamefont {Naletov}}, \bibinfo
  {author} {\bibfnamefont {N.}~\bibnamefont {Vukadinovic}}, \bibinfo {author}
  {\bibfnamefont {G.}~\bibnamefont {de~Loubens}}, \ and\ \bibinfo {author}
  {\bibfnamefont {O.}~\bibnamefont {Klein}},\ }\href {\doibase
  10.1103/PhysRevB.93.184427} {\bibfield  {journal} {\bibinfo  {journal} {Phys.
  Rev. B}\ }\textbf {\bibinfo {volume} {93}},\ \bibinfo {pages} {184427}
  (\bibinfo {year} {2016})}\BibitemShut {NoStop}%
\bibitem [{\citenamefont {Kakazei}\ \emph {et~al.}(2004)\citenamefont
  {Kakazei}, \citenamefont {Wigen}, \citenamefont {Guslienko}, \citenamefont
  {Novosad}, \citenamefont {Slavin}, \citenamefont {Golub}, \citenamefont
  {Lesnik},\ and\ \citenamefont {Otani}}]{doi:10.1063/1.1772868}%
  \BibitemOpen
  \bibfield  {author} {\bibinfo {author} {\bibfnamefont {G.~N.}\ \bibnamefont
  {Kakazei}}, \bibinfo {author} {\bibfnamefont {P.~E.}\ \bibnamefont {Wigen}},
  \bibinfo {author} {\bibfnamefont {K.~Y.}\ \bibnamefont {Guslienko}}, \bibinfo
  {author} {\bibfnamefont {V.}~\bibnamefont {Novosad}}, \bibinfo {author}
  {\bibfnamefont {A.~N.}\ \bibnamefont {Slavin}}, \bibinfo {author}
  {\bibfnamefont {V.~O.}\ \bibnamefont {Golub}}, \bibinfo {author}
  {\bibfnamefont {N.~A.}\ \bibnamefont {Lesnik}}, \ and\ \bibinfo {author}
  {\bibfnamefont {Y.}~\bibnamefont {Otani}},\ }\href {\doibase
  10.1063/1.1772868} {\bibfield  {journal} {\bibinfo  {journal} {Applied
  Physics Letters}\ }\textbf {\bibinfo {volume} {85}},\ \bibinfo {pages} {443}
  (\bibinfo {year} {2004})}\BibitemShut {NoStop}%
\bibitem [{\citenamefont {Yu}\ \emph {et~al.}(2002)\citenamefont {Yu},
  \citenamefont {Guslienko}, \citenamefont {Novosad}, \citenamefont {Otani},
  \citenamefont {Shima},\ and\ \citenamefont {Fukamichi}}]{YuCore}%
  \BibitemOpen
  \bibfield  {author} {\bibinfo {author} {\bibfnamefont {K.}~\bibnamefont
  {Yu}}, \bibinfo {author} {\bibfnamefont {K.}~\bibnamefont {Guslienko}},
  \bibinfo {author} {\bibfnamefont {V.}~\bibnamefont {Novosad}}, \bibinfo
  {author} {\bibfnamefont {Y.}~\bibnamefont {Otani}}, \bibinfo {author}
  {\bibfnamefont {H.}~\bibnamefont {Shima}}, \ and\ \bibinfo {author}
  {\bibfnamefont {K.}~\bibnamefont {Fukamichi}},\ }\href@noop {} {\bibfield
  {journal} {\bibinfo  {journal} {Phys. Rev. B}\ }\textbf {\bibinfo {volume}
  {65}} (\bibinfo {year} {2002})}\BibitemShut {NoStop}%
\bibitem [{\citenamefont {Ding}\ \emph {et~al.}(2014)\citenamefont {Ding},
  \citenamefont {N~Kakazei}, \citenamefont {Liu}, \citenamefont {Guslienko},\
  and\ \citenamefont {Adeyeye}}]{higher}%
  \BibitemOpen
  \bibfield  {author} {\bibinfo {author} {\bibfnamefont {J.}~\bibnamefont
  {Ding}}, \bibinfo {author} {\bibfnamefont {G.}~\bibnamefont {N~Kakazei}},
  \bibinfo {author} {\bibfnamefont {X.}~\bibnamefont {Liu}}, \bibinfo {author}
  {\bibfnamefont {K.}~\bibnamefont {Guslienko}}, \ and\ \bibinfo {author}
  {\bibfnamefont {A.}~\bibnamefont {Adeyeye}},\ }\href@noop {} {\bibfield
  {journal} {\bibinfo  {journal} {Scientific Reports}\ }\textbf {\bibinfo
  {volume} {4}},\ \bibinfo {pages} {4796} (\bibinfo {year} {2014})}\BibitemShut
  {NoStop}%
\bibitem [{\citenamefont {Guslienko}\ \emph {et~al.}(2015)\citenamefont
  {Guslienko}, \citenamefont {Kakazei}, \citenamefont {Ding}, \citenamefont
  {Liu},\ and\ \citenamefont {Adeyeye}}]{Guslienko2015GiantMV}%
  \BibitemOpen
  \bibfield  {author} {\bibinfo {author} {\bibfnamefont {K.~Y.}\ \bibnamefont
  {Guslienko}}, \bibinfo {author} {\bibfnamefont {G.~N.}\ \bibnamefont
  {Kakazei}}, \bibinfo {author} {\bibfnamefont {J.}~\bibnamefont {Ding}},
  \bibinfo {author} {\bibfnamefont {X.~M.}\ \bibnamefont {Liu}}, \ and\
  \bibinfo {author} {\bibfnamefont {A.~O.}\ \bibnamefont {Adeyeye}},\
  }\href@noop {} {\bibfield  {journal} {\bibinfo  {journal} {Scientific
  Reports}\ }\textbf {\bibinfo {volume} {5}},\ \bibinfo {pages} {13881}
  (\bibinfo {year} {2015})}\BibitemShut {NoStop}%
\bibitem [{\citenamefont {Eshbach}\ and\ \citenamefont
  {Damon}(1960)}]{PhysRev.118.1208}%
  \BibitemOpen
  \bibfield  {author} {\bibinfo {author} {\bibfnamefont {J.~R.}\ \bibnamefont
  {Eshbach}}\ and\ \bibinfo {author} {\bibfnamefont {R.~W.}\ \bibnamefont
  {Damon}},\ }\href {\doibase 10.1103/PhysRev.118.1208} {\bibfield  {journal}
  {\bibinfo  {journal} {Phys. Rev.}\ }\textbf {\bibinfo {volume} {118}},\
  \bibinfo {pages} {1208} (\bibinfo {year} {1960})}\BibitemShut {NoStop}%
\bibitem [{\citenamefont {Bracher}\ \emph {et~al.}(2017)\citenamefont
  {Bracher}, \citenamefont {Pirro},\ and\ \citenamefont
  {Hillebrands}}]{BRACHER20171}%
  \BibitemOpen
  \bibfield  {author} {\bibinfo {author} {\bibfnamefont {T.}~\bibnamefont
  {Bracher}}, \bibinfo {author} {\bibfnamefont {P.}~\bibnamefont {Pirro}}, \
  and\ \bibinfo {author} {\bibfnamefont {B.}~\bibnamefont {Hillebrands}},\
  }\href {\doibase https://doi.org/10.1016/j.physrep.2017.07.003} {\bibfield
  {journal} {\bibinfo  {journal} {Physics Reports}\ }\textbf {\bibinfo {volume}
  {699}},\ \bibinfo {pages} {1 } (\bibinfo {year} {2017})}\BibitemShut
  {NoStop}%
\bibitem [{\citenamefont {Kalinikos}\ and\ \citenamefont
  {Slavin}(1986)}]{Kalinikos_1986}%
  \BibitemOpen
  \bibfield  {author} {\bibinfo {author} {\bibfnamefont {B.~A.}\ \bibnamefont
  {Kalinikos}}\ and\ \bibinfo {author} {\bibfnamefont {A.~N.}\ \bibnamefont
  {Slavin}},\ }\href {\doibase 10.1088/0022-3719/19/35/014} {\bibfield
  {journal} {\bibinfo  {journal} {Journal of Physics C: Solid State Physics}\
  }\textbf {\bibinfo {volume} {19}},\ \bibinfo {pages} {7013} (\bibinfo {year}
  {1986})}\BibitemShut {NoStop}%
\bibitem [{\citenamefont {Garcia-Sanchez}\ \emph {et~al.}(2015)\citenamefont
  {Garcia-Sanchez}, \citenamefont {Borys}, \citenamefont {Soucaille},
  \citenamefont {Adam}, \citenamefont {Stamps},\ and\ \citenamefont
  {Kim}}]{PhysRevLett.114.247206}%
  \BibitemOpen
  \bibfield  {author} {\bibinfo {author} {\bibfnamefont {F.}~\bibnamefont
  {Garcia-Sanchez}}, \bibinfo {author} {\bibfnamefont {P.}~\bibnamefont
  {Borys}}, \bibinfo {author} {\bibfnamefont {R.}~\bibnamefont {Soucaille}},
  \bibinfo {author} {\bibfnamefont {J.-P.}\ \bibnamefont {Adam}}, \bibinfo
  {author} {\bibfnamefont {R.~L.}\ \bibnamefont {Stamps}}, \ and\ \bibinfo
  {author} {\bibfnamefont {J.-V.}\ \bibnamefont {Kim}},\ }\href@noop {}
  {\bibfield  {journal} {\bibinfo  {journal} {Phys. Rev. Lett.}\ }\textbf
  {\bibinfo {volume} {114}},\ \bibinfo {pages} {247206} (\bibinfo {year}
  {2015})}\BibitemShut {NoStop}%
\bibitem [{\citenamefont {Wagner}\ \emph {et~al.}(2016)\citenamefont {Wagner},
  \citenamefont {K{\'a}kay}, \citenamefont {Schultheiss}, \citenamefont
  {Henschke}, \citenamefont {Sebastian},\ and\ \citenamefont
  {Schultheiss}}]{Wagner2016MagneticDW}%
  \BibitemOpen
  \bibfield  {author} {\bibinfo {author} {\bibfnamefont {K.}~\bibnamefont
  {Wagner}}, \bibinfo {author} {\bibfnamefont {A.}~\bibnamefont {K{\'a}kay}},
  \bibinfo {author} {\bibfnamefont {K.}~\bibnamefont {Schultheiss}}, \bibinfo
  {author} {\bibfnamefont {A.}~\bibnamefont {Henschke}}, \bibinfo {author}
  {\bibfnamefont {T.}~\bibnamefont {Sebastian}}, \ and\ \bibinfo {author}
  {\bibfnamefont {H.}~\bibnamefont {Schultheiss}},\ }\href@noop {} {\bibfield
  {journal} {\bibinfo  {journal} {Nature nanotechnology}\ }\textbf {\bibinfo
  {volume} {11 5}},\ \bibinfo {pages} {432} (\bibinfo {year}
  {2016})}\BibitemShut {NoStop}%
\bibitem [{\citenamefont {Winter}(1961)}]{PhysRev.124.452}%
  \BibitemOpen
  \bibfield  {author} {\bibinfo {author} {\bibfnamefont {J.~M.}\ \bibnamefont
  {Winter}},\ }\href@noop {} {\bibfield  {journal} {\bibinfo  {journal} {Phys.
  Rev.}\ }\textbf {\bibinfo {volume} {124}},\ \bibinfo {pages} {452} (\bibinfo
  {year} {1961})}\BibitemShut {NoStop}%
\bibitem [{\citenamefont {Aliev}\ \emph {et~al.}(2011)\citenamefont {Aliev},
  \citenamefont {Awad}, \citenamefont {Dieleman}, \citenamefont {Lara},
  \citenamefont {Metlushko},\ and\ \citenamefont
  {Guslienko}}]{PhysRevB.84.144406}%
  \BibitemOpen
  \bibfield  {author} {\bibinfo {author} {\bibfnamefont {F.~G.}\ \bibnamefont
  {Aliev}}, \bibinfo {author} {\bibfnamefont {A.~A.}\ \bibnamefont {Awad}},
  \bibinfo {author} {\bibfnamefont {D.}~\bibnamefont {Dieleman}}, \bibinfo
  {author} {\bibfnamefont {A.}~\bibnamefont {Lara}}, \bibinfo {author}
  {\bibfnamefont {V.}~\bibnamefont {Metlushko}}, \ and\ \bibinfo {author}
  {\bibfnamefont {K.~Y.}\ \bibnamefont {Guslienko}},\ }\href {\doibase
  10.1103/PhysRevB.84.144406} {\bibfield  {journal} {\bibinfo  {journal} {Phys.
  Rev. B}\ }\textbf {\bibinfo {volume} {84}},\ \bibinfo {pages} {144406}
  (\bibinfo {year} {2011})}\BibitemShut {NoStop}%
\bibitem [{\citenamefont {Lara}\ \emph {et~al.}(2017)\citenamefont {Lara},
  \citenamefont {Moreno}, \citenamefont {Guslienko},\ and\ \citenamefont
  {Aliev}}]{Lara2017InformationPI}%
  \BibitemOpen
  \bibfield  {author} {\bibinfo {author} {\bibfnamefont {A.~J.}\ \bibnamefont
  {Lara}}, \bibinfo {author} {\bibfnamefont {J.~R.}\ \bibnamefont {Moreno}},
  \bibinfo {author} {\bibfnamefont {K.~Y.}\ \bibnamefont {Guslienko}}, \ and\
  \bibinfo {author} {\bibfnamefont {F.~G.}\ \bibnamefont {Aliev}},\ }\href@noop
  {} {\bibfield  {journal} {\bibinfo  {journal} {Scientific Reports}\ }\textbf
  {\bibinfo {volume} {7}},\ \bibinfo {pages} {5597} (\bibinfo {year}
  {2017})}\BibitemShut {NoStop}%
\bibitem [{\citenamefont {Landau}\ and\ \citenamefont
  {Lifshitz}(1935)}]{LANDAU199251}%
  \BibitemOpen
  \bibfield  {author} {\bibinfo {author} {\bibfnamefont {L.}~\bibnamefont
  {Landau}}\ and\ \bibinfo {author} {\bibfnamefont {E.}~\bibnamefont
  {Lifshitz}},\ }\href@noop {} {\bibfield  {journal} {\bibinfo  {journal}
  {Phys. Zeitsch. der Sow.}\ }\textbf {\bibinfo {volume} {8}},\ \bibinfo
  {pages} {153–169} (\bibinfo {year} {1935})}\BibitemShut {NoStop}%
\end{thebibliography}%

\end{document}